\documentclass[12pt]{article}

\usepackage{textcomp}
\usepackage{chetnew}
\usepackage{tikz}
\usetikzlibrary{decorations.pathmorphing}
\usetikzlibrary{decorations.markings}

\usepackage{graphicx}
\usepackage{amsfonts}
\usepackage{amsmath}
\usepackage{hyperref}
\usepackage{soul}

\newcommand{\CA}{\mathcal{A}}

\newcommand{\CH}{\mathcal{H}}

\newcommand{\CC}{\mathcal{C}}
\newcommand{\CO}{\mathcal{O}}
\newcommand{\CT}{\mathcal{T}}

\newcommand{\CN}{\mathcal{N}}

\newcommand{\CM}{\mathcal{M}}

\newcommand{\bea}{\begin{eqnarray}}
\newcommand{\eea}{\end{eqnarray}}
\newcommand{\be}{\begin{equation}}
\newcommand{\ee}{\end{equation}}

\newcommand{\la}{\label}
\newcommand{\boundellipse}[3]
{(#1) ellipse (#2 and #3)
}

\preprint{QMUL-PH-17-01,  EFI-16-29}

\title{\vspace*{-16mm} Anyonic Chains, Topological Defects, and Conformal Field Theory  \\[2mm] }

\author{Matthew Buican$^{\diamondsuit, 1}$ and Andrey Gromov$^{\clubsuit, 2}$
}
\affiliation{$^1$CRST and School of Physics and Astronomy \\ Queen Mary University of London, London E1 4NS, UK\\ \smallskip$^2$Kadanoff Center for Theoretical Physics\\ and \\Enrico Fermi Institute\\  University of Chicago, Chicago, IL 60637, USA \emails{$^{\diamondsuit}$m.buican@qmul.ac.uk, $^{\clubsuit}$gromovand@uchicago.edu}}

\abstract{Motivated by the three-dimensional topological field theory / two-dimensional conformal field theory (CFT) correspondence, we study a broad class of one-dimensional quantum mechanical models, known as anyonic chains, that can give rise to an enormously rich (and largely unexplored) space of two-dimensional critical theories in the thermodynamic limit. One remarkable feature of these systems is the appearance of non-local microscopic \lq\lq topological symmetries" that descend to topological defects of the resulting CFTs. We derive various model-independent properties of these theories and of this topological symmetry / topological defect correspondence. For example, by studying precursors of certain twist and defect fields directly in the anyonic chains, we argue that (under mild assumptions) the two-dimensional CFTs correspond to particular modular invariants with respect to their maximal chiral algebras and that the topological defects descending from topological symmetries commute with these maximal chiral algebras. Using this map, we apply properties of defect Hilbert spaces to show how topological symmetries give a handle on the set of allowed relevant deformations of these theories. Throughout, we give a unified perspective that treats the constraints from discrete symmetries on the same footing as the constraints from topological ones.
}

\date{January 2017}

\begin{document}

\setcounter{tocdepth}{2}

\maketitle
\toc

\section{Introduction}
Quantum systems in three space-time dimensions allow for a well-known generalization of the usual bosonic / fermionic exchange statistics of identical particles called ``anyonic" statistics\cite{leinaas1977theory,wilczek1982quantum}. When anyons are interchanged, the resulting quantum state may pick up exotic phases different from $0$ and $\pi$ (the abelian case) or may be acted upon by non-commuting unitary matrices (the non-abelian case).

Anyons have deep connections to many areas of current interest in theoretical physics. One of their simplest non-abelian incarnations, the so-called \lq\lq Fibonacci" anyons, can be harnessed to construct models of universal quantum computation (see \cite{wang2010topological} and references therein).\footnote{In the language of high-energy theoretical physics, one can think of these anyons as arising in $G_2$ Chern-Simons theory at level one.} More generally, anyons are excitations of theories exhibiting topological order (i.e., gapped systems lacking a local order parameter) like discrete gauge theories and systems that exhibit the fractional quantum Hall effect (FQHE).

Apart from being interesting experimental realizations of strongly coupled physics, FQHE models are thought to be real-world examples \cite{moore1991nonabelions} of the three-dimensional topological field theory (TFT) / two-dimensional conformal field theory (CFT) correspondence originally discussed in \cite{Witten:1988hf,Moore:1989yh,Moore:1989vd}. Under this correspondence, the FQHE groundstates are described by conformal blocks of boundary CFTs, and anyonic particles  are described by chiral vertex operators of these boundary theories.

In this paper, we study a vast set of models that exhibit anyonic relations between TFTs and CFTs and can potentially be used to better understand the above systems (as well as to explore the space of two-dimensional CFTs). Our starting point is a generalization of certain quantum mechanical models called \lq\lq anyonic chains" \cite{feiguin2007interacting}. The basic idea is to treat the anyons as elements of an input fusion category, $\CC_{\rm in}$, that satisfy a multiplication rule generalizing the usual multiplication of spins. One then constructs fusion trees by specifying a lattice of external anyonic states of length $L$ while allowing the labels of internal links (i.e., the interactions) to take on any values in $\CC_{\rm in}$ consistent with the fusion rules. The resulting Hilbert spaces are somewhat reminiscent of those encountered in the construction of simple spin chains like the Heisenberg model, but, in general, the spaces of anyonic states do not factorize into tensor products of local Hilbert spaces.

In order to move beyond topology and introduce a notion of energy that allows us to make closer contact with CFT, we will specify a Hamiltonian, $H$, for the anyonic system \cite{feiguin2007interacting}. Quite remarkably, in the thermodynamic limit (i.e., in the limit that $L\to\infty$), one often finds an effective two-dimensional CFT description of the one-dimensional chain (which we will refer to as the \lq\lq output" CFT). While similar behavior is known to occur in spin chains like the Heisenberg model, the mechanism that gives rise to this critical effective behavior in anyonic chains is somewhat different. Indeed, we will see that certain types of non-local operators (also known as \lq\lq topological symmetries,"), $Y_{\ell}$, play a starring role \cite{feiguin2007interacting, gils2013anyonic}. Roughly speaking, the $Y_{\ell}$ are in one-to-one correspondence with the simple elements of $\CC_{\rm in}$, and they can be used  to rule out relevant terms in the effective action if and only if these deformations are in \lq\lq topologically non-trivial" sectors of the theory. As a result, if there are sufficiently many of these symmetries, we can end up with a critical effective theory that has no relevant perturbations invariant under the topological symmetries.\footnote{Otherwise, we may end up with a gapped theory.}

Given this construction, it is natural to ask which objects in the output CFT the topological symmetries correspond to. As we will see, the $Y_{\ell}$ descend to topological defects \cite{Petkova:2000ip} of the effective theory. Basic aspects of this correspondence appeared in \cite{pfeifer2012translation}. These connections were then more generally and explicitly formulated in \cite{Aasen:2016dop}. In particular, a connection between the topological symmetries and the topological defects of \cite{Petkova:2000ip} was first discussed in \cite{Aasen:2016dop}. A lattice realization of topological defects was also given in \cite{Hauru:2015abi}. One aspect of our work below will be to find additional evidence for this correspondence. Moreover, since topological defects are rich objects in their own right---with interesting Hilbert spaces of twist, defect, and junction fields localized on them---we are able to gain new perspectives on the topological symmetries themselves and the role they play in giving rise to critical behavior.

In what follows, we eschew a detailed study of specific examples of anyonic chains in order to focus on general properties of this topological symmetry / topological defect correspondence. Our main motivations for this approach are a desire to better understand the role topological defects play in general FQHE and condensed matter systems, and to determine to what extent anyonic chains can be used to further explore the space of two-dimensional CFTs (perhaps as an alternative to more conventional conformal bootstrap techniques).

Implementing this general approach, we find the following model-independent results
\begin{itemize}
\item The topological symmetries, $Y_{\ell}$, realize the input fusion algebra, $\CA_{\rm in}$, of $\CC_{\rm in}$ (which we will take to be modular), while---up to some important caveats we discuss in Section \ref{correspondence}---the corresponding topological defects, $D_{\ell}$, in the output CFT, $\CT_{\rm out}$, realize a fusion sub-algebra, $\CA'$, of the output fusion algebra, $\CA_{\rm out}$, that is isomorphic to $\CA_{\rm in}$, i.e.,\footnote{A similar result is independently obtained in the upcoming work \cite{Aasen2}. We thank the authors of \cite{Aasen2} for sharing their work with us prior to publication.}
\begin{equation}\label{fusionalg}
\CA_{\rm in}\simeq\CA'\subseteq\CA_{\rm out}~.
\end{equation}
Moreover, topological symmetry operators have the same eigenvalues as the corresponding topological defects
\begin{equation}\label{equaleval}
Y_{\ell}|m,\overrightarrow\alpha\rangle=\lambda_{\ell,m}|m,\overrightarrow\alpha\rangle\ \ \ \Leftrightarrow\ \ \ D_{\ell}|\Phi_{m,\overrightarrow A}\rangle=\lambda_{\ell, m}|\Phi_{m,\overrightarrow A}\rangle~,
\end{equation}
where $|m,\overrightarrow\alpha\rangle$ is a state in the Hilbert space of the anyonic chain that has topological quantum numbers corresponding to an element $\varphi_m\in\CC_{\rm in}$ (the set of quantum numbers, \lq\lq$\overrightarrow\alpha$," distinguishes between the many states that have the same topological quantum numbers), and $|\Phi_{m,\overrightarrow A}\rangle$ is a state in $\CT_{\rm out}$ that descends from $|m,\overrightarrow\alpha\rangle$ (note that all CFT states can be written in this way\footnote{Although, in general, only certain $\Phi_{m,\overrightarrow A}$ CFT operators \lq\lq directly descend" from $\varphi_m$ in the sense of realizing the corresponding fusion rules in $\CA'$.}).

\item If $\lambda_{\ell,0}$ is the eigenvalue of the ground state, $|0\rangle$, under the action of $D_{\ell}$ (or, equivalently, of $Y_{\ell}$), we have
\begin{equation}\label{evalcomm}
 \lambda_{\ell,m}=\lambda_{\ell,0}~, \ \ \ \forall\ell\in\CA' \ \ \ \Leftrightarrow\ \ \ \left[D_{\ell},\Phi_{m,\overrightarrow A}\right]=0~, \ \ \ \forall\ell\in\CA'~,
\end{equation}
where $\Phi_{m,\overrightarrow A}$ is the CFT operator corresponding to the state $|\Phi_{m,\overrightarrow A}\rangle$. Note that \eqref{evalcomm} is the CFT version of the topological protection mechanism discussed in \cite{feiguin2007interacting,gils2013anyonic}. In particular, the vanishing of the microscopic commutators
\begin{equation}\label{microcomm}
\left[Y_{\ell},H\right]=0~,\ \ \ \forall\ell\in\CA_{\rm in}~,
\end{equation}
ensures that if a putative deformation of the effective theory, $\delta H_{\rm eff} = \lambda \oint\Phi_{m,\overrightarrow A}$, does not satisfy \eqref{evalcomm}, then it cannot be generated upon coarse graining.

\item The \lq\lq$\Leftarrow$" direction of \eqref{evalcomm} is a straightforward application of the CFT state-operator correspondance as well as of basic properties of fusion algebras and topological defects (see Theorem \ref{Thm1} below). On the other hand, the \lq\lq$\Rightarrow$" direction is non-trivial. Indeed, we first prove a partial converse of Theorem \ref{Thm1}, that depends on properties of defect Hilbert spaces, in Theorem \ref{Thm2}. Then, by combining the quantum mechanical constraint in \eqref{equaleval} with unitarity (in the guise of Theorem \ref{Thm3}) we complete the proof of \eqref{evalcomm} (in Theorem \ref{Thm4}).

\item Our proof of the \lq\lq$\Rightarrow$" direction of \eqref{evalcomm} uses unitarity (see Theorem \ref{Thm3}). Therefore, it is possible that there are some non-unitary anyonic chains (see \cite{ardonne2011microscopic} for basic examples of non-unitary chains) that have $\left[D_{\ell}, \Phi_{m,\overrightarrow A}\right]\ne0$ even though $\lambda_{\ell, m}=\lambda_{\ell,0}$ $\forall\ell\in\CA'$. If such examples do exist, then these non-unitary chains may exhibit an enhanced topological protection of critical behavior in the thermodynamic limit (since relevant deformations in the topological sector of the identity might still be ruled out).

\item Under relatively mild assumptions, the subset of topological defects of $\CT_{\rm out}$ that descend from topological symmetries of the anyonic chain commute with the full left and right chiral algebras of $\CT_{\rm out}$. For example, if the output theory is the 3-state Potts model, then these topological defects not only commute with the left and right Virasoro algebras but also commute with the full left and right $W_3$ algebras.

\item We find quantum mechanical ancestors of some of the twist and defect fields of the $D_{\ell}$ in the anyonic chain itself.\footnote{Similar results are independently obtained in the upcoming article \cite{Aasen2}.} As we will see, the spectrum of these ancestors shows that the anyonic chain contains seeds of the modular properties of $\CT_{\rm out}$ even though $\CC_{\rm in}$ itself need not be modular in general (although, to prove the \lq\lq$\Rightarrow$" direction of \eqref{evalcomm}, we will assume modularity of $\CC_{\rm in}$).

\item Discrete symmetries can be treated on a similar footing. Operators generating these symmetries correspond to group-like defects that have group-like fusion rules. Discrete symmetries of the microscopic theory can be enforced macroscopically via commutators similar to those in \eqref{evalcomm}. These symmetries are often crucial for ensuring critical behavior in the $L\to\infty$ limit (see the examples in Section \ref{Exapp}).

\item In order to set the stage for studying the potential implications of our work on the space of two-dimensional CFTs, we find simple constraints on the possible $\CC_{\rm in}$ that can give rise to a particular output unitary modular tensor category (UMTC), $\CC_{\rm out}$ (where we assume that $\CC_{\rm out}$ is the UMTC of some rational CFT, $\CT_{\rm out}$). As an illustrative example of these constraints, we note that (up to the caveats of Section \ref{correspondence}) if $\CC_{\rm out}=D(S_3)$, where $D(S_3)$ is the quantum double of $S_3$, then the only possible input modular fusion category is $\CC_{\rm in}=D(S_3)$ as well.\footnote{\label{isingfoot} A similar statement holds for the Ising model fusion category. Note that in making this statement, we do not distinguish between input data that differs by Frobenius-Schur indicators \cite{rowell2009classification}, so the input fusion category can be $\CC_{\rm in}={\rm Rep}\left[su(2)_2\right]$ (where ${\rm Rep}\left[g\right]$ are the representations of the algebra $g$). In fact, such a case arises in \cite{gils2013anyonic}.}
\end{itemize}

Let us now briefly summarize the plan of this paper. In the next section, we introduce anyonic chains in much greater detail. In order to motivate the various possible interactions present in these theories, we remind the reader of analogous interactions in the Heisenberg spin chain. We then discuss the various topological symmetries of anyonic chains and explain how they differ from the global symmetries of the Heisenberg chain. We move on to study the thermodynamic limit of the anyonic chains and the effective CFT description while placing special emphasis on further explaining the map between topological symmetries and topological defects. In the following section, we discuss the more detailed properties of topological defects, including how to compute the Hilbert spaces of fields localized on defects. By relating some of these defect fields to precursors in the anyonic chains, we argue that the defects descending from topological symmetries commute with the left and right maximal chiral algebras of $\CT_{\rm out}$. We then use facts about defect Hilbert spaces combined with constraints from unitarity and the quantum mechanical inputs of our theories to prove \eqref{evalcomm}. We conclude with a brief discussion of examples and applications as well as a list of open problems and future directions.

\section{Anyonic Chains}
\la{chains}

In order to introduce basic aspects of the correspondence between 1D quantum mechanical systems and 2D CFTs, we begin this section by discussing the Heisenberg spin-${1\over2}$ chain. After briefly mentioning its generalizations, we move on to discuss anyonic chains \cite{feiguin2007interacting, gils2009collective,trebst2008short, trebst2008collective, gils2013anyonic, ardonne2011microscopic}, and we review the construction of the resulting Hilbert spaces, Hamiltonians, and topological symmetries. 

However, several words of caution are in order: while the Heisenberg chain allows for various interactions that are closely related to those found in anyonic chains, these latter chains are not simple generalizations of the spin-${1\over2}$ Heisenberg chain. Instead, they are qualitatively different in several respects. For example, anyonic chains have non-local symmetries that are in one-to-one correspondence with elements of the fusion algebra that the chain is built from. On the other hand, the spin-${1\over2}$ Heisenberg model has global symmetries that are in one-to-one correspondence with its symmetry group elements. These facts have profound implications for the resulting effective CFTs (when they exist): in the anyonic chain case, the non-local symmetries are related to defects that (under relatively mild assumptions) commute with the full chiral algebra. On the other hand, in the spin-${1\over2}$ Heisenberg case, the global symmetries are related to defects that do not commute with the full chiral algebra (and, moreover, are defined in terms of exponentials of CFT Noether charges).

Another important difference between the Heisenberg chain and the anyonic chains is that the Hilbert space of the Heisenberg chain factorizes while the Hilbert space of the typical anyonic chain does not (this statement is, in turn, related to a richer structure of quantum dimensions in the anyonic case). In section \ref{su2kexample}, we describe various explicit examples of anyonic chains and comment on the extent to which one can view these models as deformations of the Heisenberg spin-${1\over2}$ chain.

\subsection{The Heisenberg spin-$\frac{1}{2}$ chain and its generalizations}

\begin{figure}
\centerline{\includegraphics[scale=0.5]{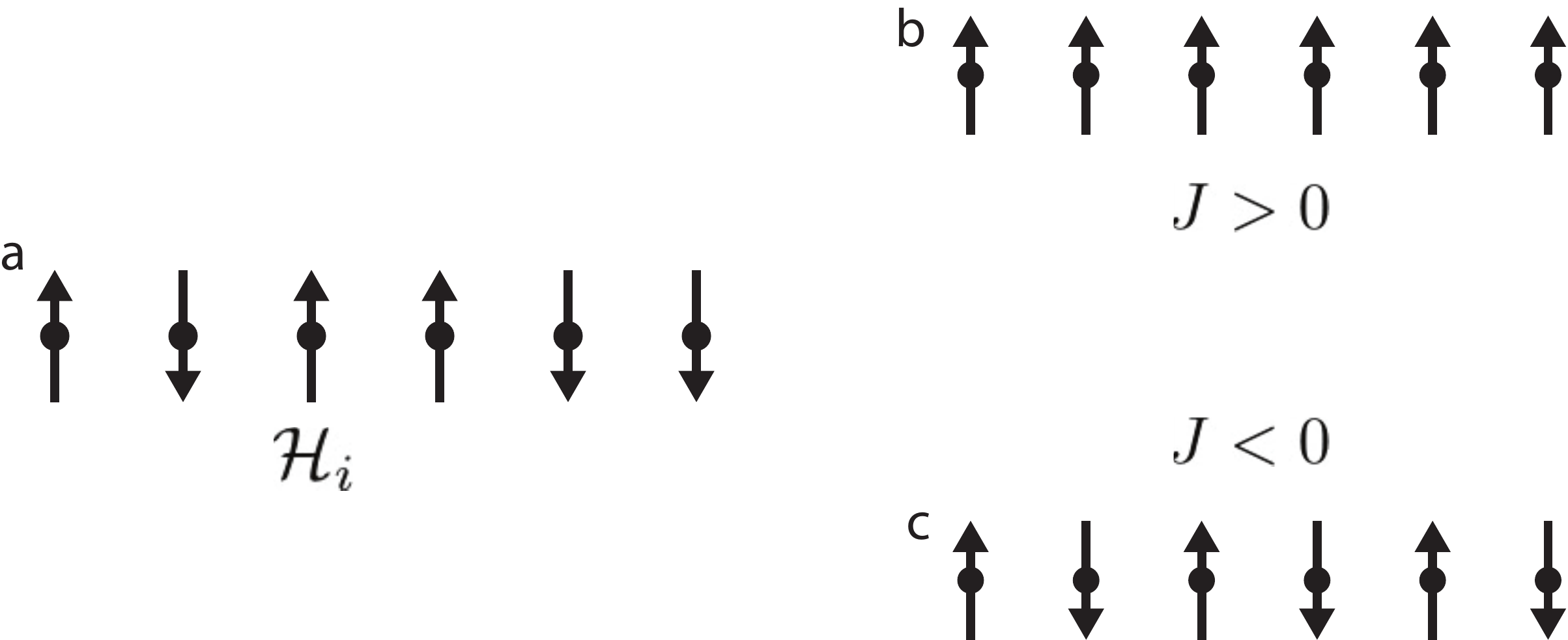}}
\caption{\textbf{(a)} The usual graphical representation of the Heisenberg chain. Each arrow indicates a state in the local Hilbert space, $\CH_i$. ~\textbf{(b)} The graphical representation of the ferromagnetic ground state. Every spin pair forms a spin-$1$ representation.~\textbf{(c)} The graphical representation of the anti-ferromagnetic groundstate. Every spin pair forms a spin-$\frac{1}{2}$ representation.}
\label{Hilbert}
\end{figure}

The Heisenberg spin chain is an integrable model of a 1D magnet \cite{fradkin2013field}. Let us first define the Hilbert space: we take a 1D lattice and attach to every node of the lattice a local Hilbert space, $\CH_i \approx \mathbb C$. The total Hilbert space is then the tensor product of the local Hilbert spaces
\be\la{HeisenbergHil}
\CH = \CH_1 \otimes \CH_2 \otimes \cdots \otimes \CH_L~,
\ee
where $L$ is the number of nodes (Fig. \ref{Hilbert}). The traditional Heisenberg Hamiltonian is given by a sum of local terms
\be\la{HeisenbergH}
H = -J \sum_{i=1}^L \vec S_i \cdot \vec S_{i+1}~,
\ee

\noindent
where the spin operator, $S^a_i$, acts as $S^a_i =  \frac{1}{2} \sigma^a$ on $\CH_i$ and as the identity everywhere else. The commutation relations of spin operators are the usual ones for $su(2)$ at each site, and the Hamiltonian in \eqref{HeisenbergH} has a global $su(2)$ symmetry.

While the absolute value of the coupling constant has no physical significance and can be rescaled away, the sign of $J$ plays an important role. To illustrate this point (and for future reference when discussing anyonic chains) we will rewrite the Hamiltonian in a more intuitive form. To that end, consider the quantity $\vec S^{\rm tot}_i = \vec S_i + \vec S_{i+1}$. The magnitude $|\vec S^{\rm tot}_i|^2$ can either be equal to $0$ or $1$ (i.e., in the language we will use below, the fusion of two spins has two channels), which follows from representation theory of $su(2)$. It is a matter of elementary algebra to check that $|\vec S^{\rm tot}_i|^2 = 0\cdot\Pi_i^{(0)}+ 2\cdot\Pi_i^{(1)}$, where $\Pi_i^{(s)}$ is a projector onto spin-$s$ states, and the coefficient is given by $s(s+1)$. Thus, the local term $ \vec S_i \cdot \vec S_{i+1}$ can be written as
\be
 \vec S_i \cdot \vec S_{i+1} = -\Pi_i^{(0)} + \frac{1}{4} \mathbb I_i~,
\ee
and the Hamiltonian \eqref{HeisenbergH} can therefore be re-written (up to a constant) as 
\be\la{Heisenbergprojector}
H = J\sum_{i=1}^L \Pi_i^{(0)}~.
\ee
Now the physical meaning of $J$ should be very clear. If $J>0$, then the neighboring spins prefer to align and form spin-$1$ combinations, so that $\Pi_i^{(0)}$ acting on each pair is $0$ (as in Fig. \ref{Hilbert} {\bf(b)}). This scenario is called the {\it ferromagnetic} (F) case. When $J<0$, spins prefer to maximize $\Pi_i^{(0)}$ by aligning in opposite directions and forming spin-$0$ singlets (as in Fig. \ref{Hilbert} {\bf(c)}). This situation is known as the {\it anti-ferromagnetic} (AF) case. When we introduce anyonic chains, we will write the Hamiltonian in terms of certain projectors onto fusion channels conceptually similar to those appearing in \eqref{Heisenbergprojector}.

In the thermodynamic limit, $L \rightarrow \infty$, the chain becomes critical (i.e., there is no energy gap between the excitations). When $J<0$, this theory is the compact free boson CFT at the self-dual radius, which is just the $su(2)_1$ Wess-Zumino-Witten (WZW) theory \cite{fradkin2013field}. The reason for the appearance of the $su(2)$ symmetry group is not accidental and is due to the presence of the global $su(2)$ of the microscopic Hamiltonian \eqref{HeisenbergH}. When $J>0$, the situation is dramatically different. The theory remains gapless, but it has dynamical critical exponent $z=2$, and, consequently, is scale invariant, but not Lorentz invariant.

We want to emphasize that in the AF case, the global $su(2)$ symmetry can be seen in the thermodynamic limit as a set of charges with corresponding Noether currents (which, in turn, appear in the spectrum of local CFT operators). Moreover, these symmetries map to the only topological defects appearing in the theory: exponentials of the corresponding Noether charges \cite{Fuchs:2007tx}.\footnote{See \cite{Gaiotto:2014kfa} for a recent discussion of the topological nature of charges.} Since these charges are non-abelian, they do not commute with the full affine-Kac-Moody (AKM) chiral algebra of the output CFT. We will see, however, that the situation is quite different in the case of anyonic chains.

Before concluding the introduction, let us note that the Heisenberg model also admits many generalizations, the most famous of which is to allow for spin-$s$ operators. In this case, the Hamiltonian can take a more complicated form. For example, for spin-$1$, we can have
\be\label{Hgen}
H^{(1)} = \sum_{i=1}^N \Big(\cos\theta \,\vec S_i \cdot\vec S_{i+1} + \sin\theta \,  (\vec S_i \cdot \vec S_{i+1})^2\Big)~.
\ee
The phase diagram of this model is quite complicated (see \cite{gils2013anyonic} for an overview). One important handle on the behavior of the chain is a general result (due to Haldane \cite{haldane1983nonlinear}) which states that in the AF Heisenberg case, if the spin $s$ is half-integer, then the spin chain is gapless (and conformal). On the other hand, if $s$ is integer, the Heisenberg spin chain is gapped (although, for example, one finds various gapless phases by moving away from the Heisenberg point and adjusting $\theta$ in \eqref{Hgen}). In the gapless case, the critical theory is the $su(2)_{1}$ WZW theory \cite{affleck1987critical}. In the gapped regime, the Lieb-Schultz-Mattis theorem \cite{lieb1961two} guarantees that the ground state is degenerate.

Another possible generalization is the inclusion of local, but not ``nearest neighbor," couplings (so-called \lq\lq Majumdar-Ghosh" couplings \cite{majumdar1969next}) of the form
\be
\delta H_k = J_k \sum_{i=1}^L \vec S_{i}\cdot \vec S_{i+k}~.
\ee
These couplings can dramatically change the behavior of the chain in the thermodynamic limit. For example, the AF spin-$\frac{1}{2}$ chain becomes gapped when $J_2 = \frac{1}{2}J_1$ and all other $J_k$ vanish \cite{majumdar1969next}.

In general, one can construct an arbitrarily complicated Hamiltonian by combining higher spin with the interaction of $k$-nearest neighbors. In this case, the full phase diagram is unknown; however, it was recently argued that in the AF case, no matter how complicated the interactions are, if the microscopic Hamiltonian is translation invariant and the chain is gapless in the thermodynamic limit then it is generically described by an $su(2)_1$ WZW theory  \cite{furuya2015symmetry}.

In the next subsection, we will introduce anyonic chains and interactions that are similar in spirit to the arbitrarily complicated ones we introduced in this section. Consequently, it is interesting to characterize the critical theories that can appear in the thermodynamic limit.

\subsection{Anyonic chains}
In a certain limited sense, anyonic chains are generalizations of the Heisenberg model, where interacting spins are replaced by interacting anyons \cite{feiguin2007interacting} (see above for some caveats to this statement). To define these models, we need a fusion category, $\CC_{in}$ (see \cite{kitaev2006anyons,wang2010topological} for more detailed reviews). For our purposes below, we merely note that $\CC_{\rm in}$ consists of a set of simple objects\footnote{These are the objects that we can use to build all the other objects in $\CC_{\rm in}$.} satisfying a fusion algebra, $\CA_{\rm in}$, that generalizes the multiplication of spins. We will also use the fact that $\CC_{\rm in}$ contains a set of isomorphisms, called \lq\lq$F$-symbols", that allow us to move between different fusion channels of the objects in an associative way. The case $\CC_{in} = {\rm Rep}\left[su(2)_k\right]$ was studied in  \cite{feiguin2007interacting, gils2009collective,trebst2008short, trebst2008collective, gils2013anyonic, ardonne2011microscopic}, but we will be more general.

\begin{figure}
\centerline{\includegraphics[scale=1]{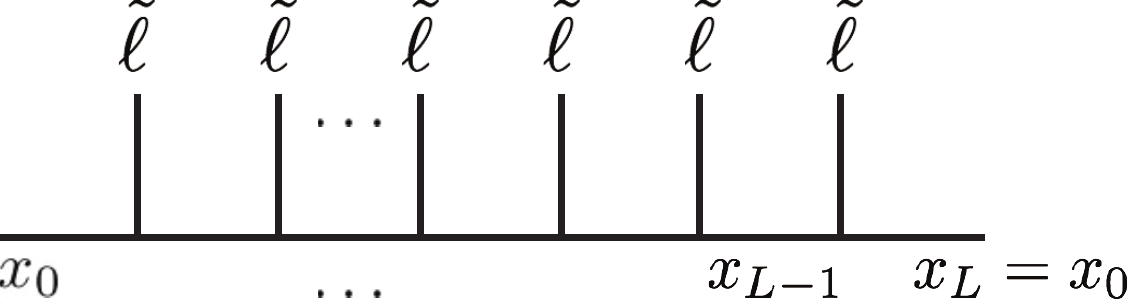}}
\caption{Fusion trees label states in the Hilbert space. For every collection of $\{x_i\}$ allowed by the fusion rules, there is a state in the Hilbert space. This chain wraps a topologically non-trivial cycle.} 
\label{FD}
\end{figure}

To proceed with our construction, let $\{ \ell_1,\ell_2, \ell_3, \cdots \}$ be the simple objects in $\mathcal C_{\rm in}$ satisfying the fusion algebra $\CA_{\rm in}$
\be
\ell_1 \times \ell_2 = \sum_{\ell_3} N^{\ell_3}_{\ell_1\ell_2} \ell_3~,
\ee
where the positive integers, $N^{\ell_3}_{\ell_1\ell_2}$, are the fusion coefficients. Given this data, we can describe the Hilbert space of the problem. First, fix an integer $L$, and take $L$ objects $\{\tilde\ell_1, \tilde\ell_2,\cdots, \tilde\ell_L\} \in \mathcal \CC_{in}$. Next, consider a periodic fusion diagram (wrapping a non-contractible cycle in the ambient spacetime) as in Fig \ref{FD}. Every admissible fusion diagram corresponds to a state in the Hilbert space, $| x_0, x_1,\ldots, x_{L-1}\rangle$. The Hamiltonian---as well as the other operators---act on the Hilbert space of admissible fusion trees. In what follows, we will also assume translational invariance: 
\begin{equation}\label{tinv}
\tilde\ell_1=\tilde\ell_2=\cdots=\tilde\ell_L=\tilde\ell~.
\end{equation}

It is crucial to note that this Hilbert space does not generally have the direct product structure of \eqref{HeisenbergHil}. For example, in the case of the ${\rm Rep}\left[su(2)_k\right]$ chain with $k$ finite, the Hilbert space is actually smaller (in the limit $L\gg1$) than the Hilbert space of the spin-$\tilde\ell$ chain since this scaling is controlled by the quantum dimension of $\tilde\ell$ (which is strictly smaller than $2\tilde\ell+1$) \cite{gils2013anyonic}.

Next, following \cite{feiguin2007interacting}, let us define the Hamiltonian by studying the possible fusion outcomes of the external anyons
\be
\tilde\ell \times \tilde\ell = \sum_{\ell'} N^{\ell'}_{\tilde\ell\tilde\ell} \ell'~.
\ee
In analogy with the Heisenberg chain, we introduce a projector, $P^{(\tilde\ell)(\ell')}_i$, that allows us to assign higher or lower energy for the fusion of neighboring anyons into channel $\ell'$
 \be\la{Hanyon}
H =  J\sum_{i}P^{(\tilde\ell)(\ell')}_i~.
 \ee
To derive an explicit expression for this projector, we need the $F$-symbols mentioned above.

\begin{figure}
\centerline{\includegraphics[scale=0.8]{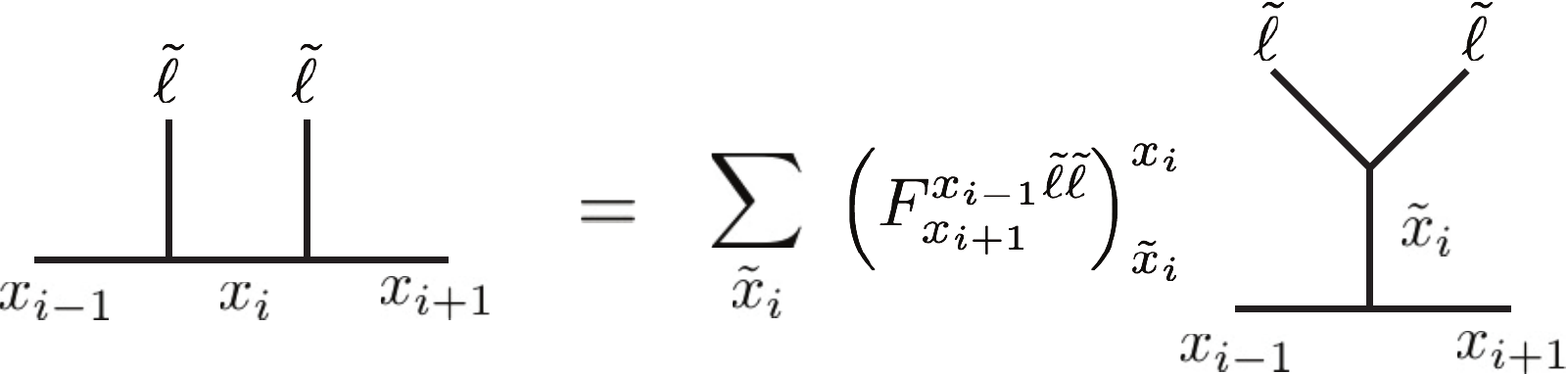}}
\caption{It is more convenient to define the action of the Hamiltonian in a different basis, related by an $F-$move to the original one. } 
\label{Hamiltonian1}
\end{figure}

\begin{figure}
\centerline{\includegraphics[scale=0.8]{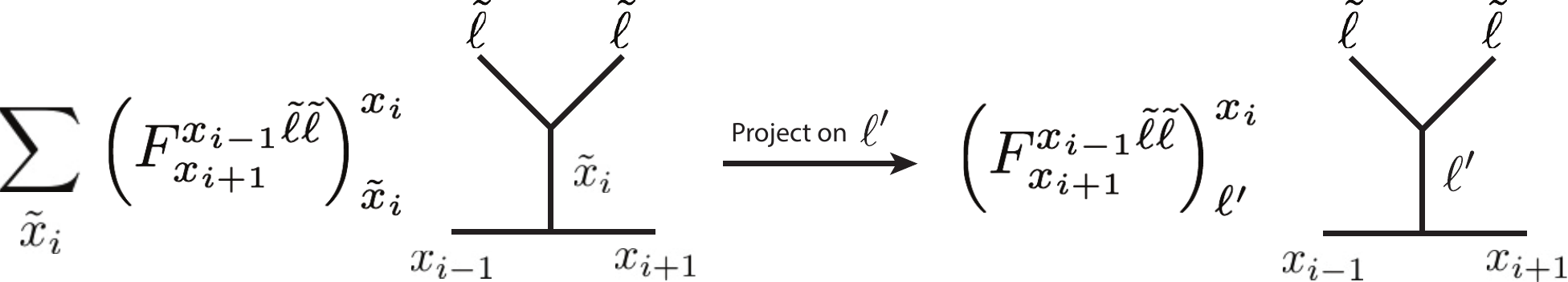}}
\caption{We choose the projector that forces the fusion outcome of nearest neighbors to be equal to $\ell^\prime$. } 
\label{Hamiltonian2}
\end{figure}

\begin{figure}
\centerline{\includegraphics[scale=0.7]{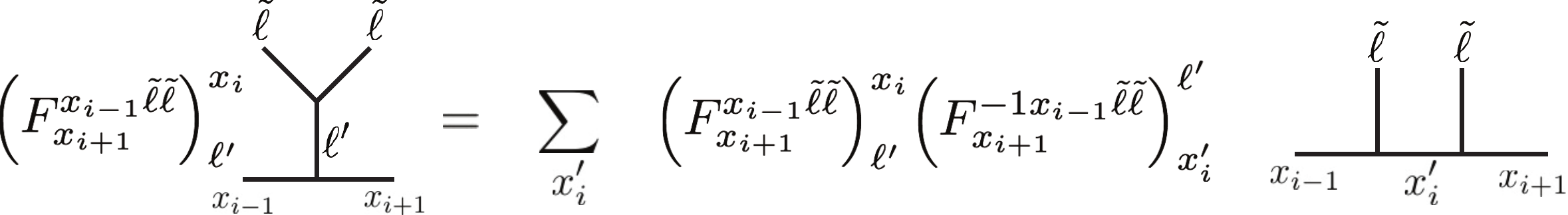}}
\caption{An $F$-move back to the original basis gives an explicit formula for the Hamiltonian.} 
\label{Hamiltonian3}
\end{figure}

Consider a fusion diagram with two external legs as in Fig \ref{Hamiltonian1}. First, we switch fusion chanels via the application of an $F$-symbol. In the new basis, the fusion diagram represents the fusion of two $\tilde\ell$ anyons with outcome $\tilde x_i$. The projector, $P^{(\tilde\ell)(\ell')}_i$, is defined in such a way that the fusion outcome $\ell'$ is energetically favored or disfavored depending on the sign of $J$ (corresponding to the F case or the AF case).
Next, consider Fig \ref{Hamiltonian2}. We apply the projector on the fusion channel $\tilde x_i=\ell'$. Finally, we make the transformation back to the original basis using the inverse $F-$symbol as in Fig \ref{Hamiltonian3}.
Thus, the action on a state $|x_{i-1},x_{i},x_{i+1}\rangle$ is 
\be
P^{(\tilde\ell)(\ell')}_i |x_{i-1},x_{i},x_{i+1}\rangle = \sum_{x^\prime_i} \left(F^{x_{i-1}\tilde\ell\tilde\ell}_{x_{i+1}}\right)_{\ell'}^{x_i} \left(F^{-1}{}^{x_{i-1}\tilde\ell\tilde\ell}_{x_{i+1}}\right)_{x^\prime_i}^{\ell'} |x_{i-1}, x^\prime_i, x_{i+1}\rangle~,
\ee
and the Hamiltonian is given by \eqref{Hanyon}.

Several comments are in order. First, the name ``anyon'' usually indicates extra braiding structure in addition to fusion. Strictly speaking, all we really needed so far is the fusion structure.\footnote{In order for these models to capture aspects of FQHE physics, we need to assume a braiding structure.} Thus, it would be more appropriate to call the above construction a ``fusion chain." Second, not at any point did we need to enforce unitarity (as noted in the introduction and as will be discussed further below, some non-unitary chains may exhibit stronger topological protection of gaplessness). Indeed, non-unitary models have been considered previously \cite{ardonne2011microscopic}. The models in\cite{ardonne2011microscopic} are described by non-unitary $F$-symbols with non-Hermitian Hamiltonians, and the critical theories are non-unitary RCFTs. Finally, just as in the Heisenberg case, the Hamiltonians we study can be made more complicated by introducing $k$-nearest neighbor interactions or terms that correspond to the fusion of several anyons \cite{trebst2008short}. In order to understand the general structure of the thermodynamic limit, it is then useful to understand the essential features that must be present in any emergent critical theory.

\subsection{Non-local symmetries}

To get a handle on the emergent theory, it turns out to be useful to understand the role played by certain non-local operators that commute with the anyonic chain Hamiltonian \cite{feiguin2007interacting,gils2013anyonic}. More precisely, for every object $\ell\in \CC_{in}$, there exists a non-local operator, $Y_{\ell}$, that commutes with the Hamiltonian (although, as we will discuss further below, some of the $Y_{\ell}$ may not be independent). These operators are referred to as ``topological symmetries'' in \cite{feiguin2007interacting,gils2013anyonic}.

\begin{figure}
\centerline{\includegraphics[width=\linewidth]{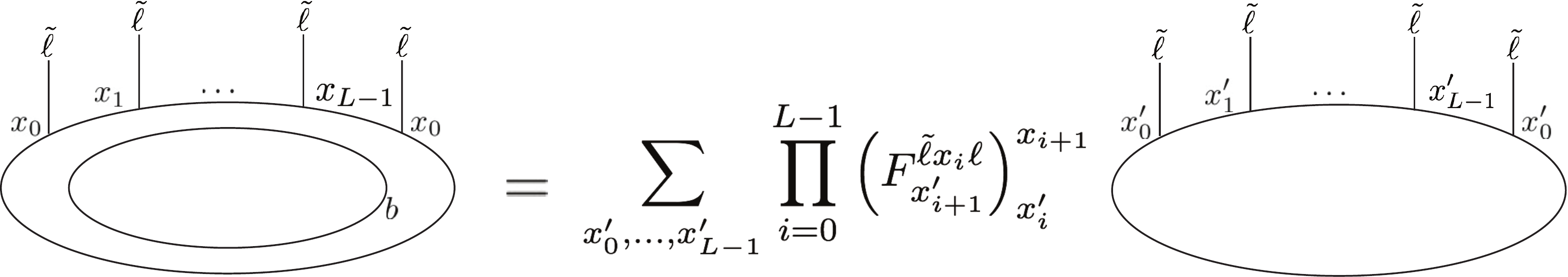}}
\caption{The action of the topological symmetry. The circles labeled by the $x_i$ wrap topologically non-trivial cycles.}
\label{TopSym}
\end{figure}

To define such an operator, we consider a process of fusing an anyonic line, $\ell$, into the chain as shown in Fig \ref{TopSym}.
 The matrix elements of the $Y_{\ell}$ operator are given by \cite{feiguin2007interacting} 
 \be
 \langle x_0^\prime,x_1^\prime,\cdots,x_{L-1}^\prime | Y_{\ell} |x_0,x_1,\cdots,x_{L-1}\rangle = \prod_{i=0}^{L-1} \left(F^{\tilde\ell x_{i}\ell}_{x_{i+1}^\prime}\right)_{x^\prime_{i}}^{x_{i+1}}~.
\ee
All operators $Y_{\ell}$ can be diagonalized simultaneously with the Hamiltonian \eqref{Hanyon}. We will denote the eigenvalues of $Y_{\ell}$ as $\lambda_{\ell,m}$ and refer to these eigenvalues as ``topological quantum numbers.'' 

\begin{figure}
\centerline{\includegraphics[width=\linewidth]{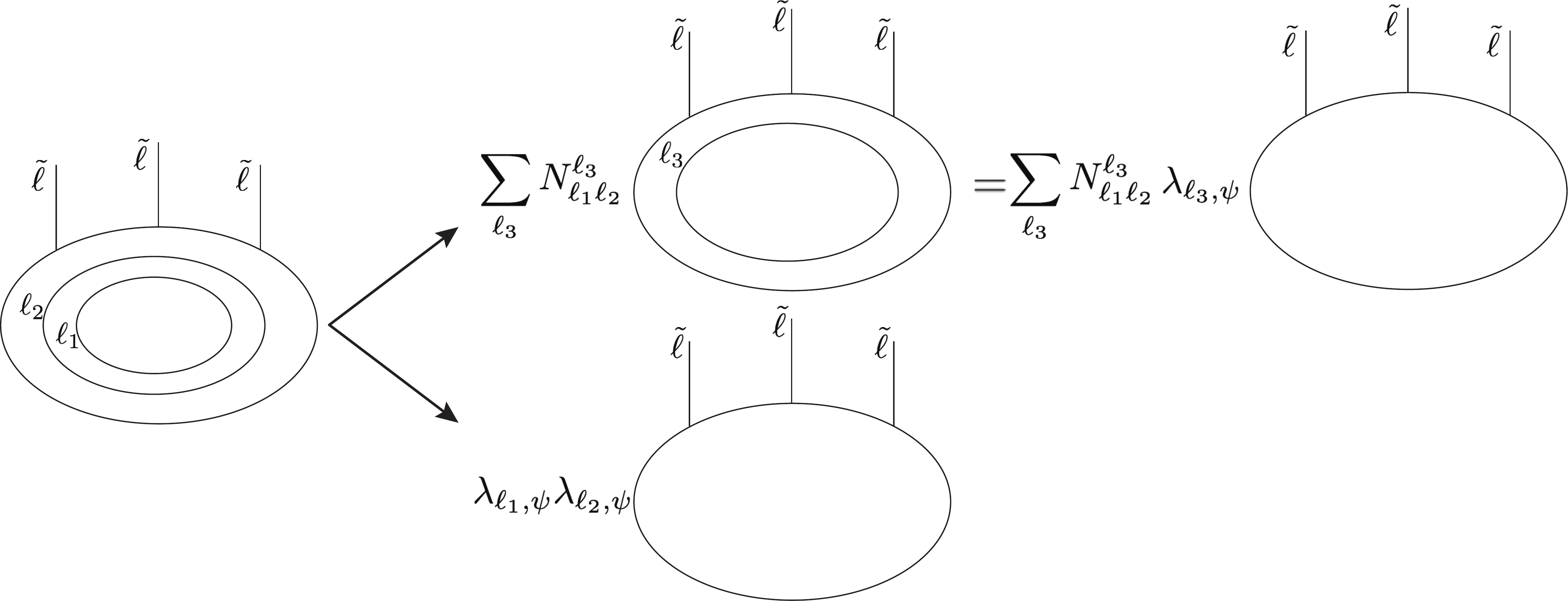}}
\caption{We show that eigenvalues of the topological symmetry operators satisfy the fusion rules as polynomial equations. We pick an eigenstate of a Hamiltonian and nucleate a pair of anyons $\ell_1$ and $\ell_2$ inside the spine of the fusion diagram. There are two ways to evaluate this diagram. First, we fuse $\ell_1$ and $\ell_2$ to get the sum $\sum_{\ell_3} N^{\ell_3}_{\ell_1\ell_2} \ell_3$. Then we fuse $\ell_3$ into the chain. Since we picked an eigenstate of the Hamiltonian and topological symmetry operators, fusing $\ell_3$ leads to multiplication by the eigenvalue $\lambda_{\ell_3,\psi}$. Alternatively, we could have fused first $\ell_1$ and then $\ell_2$ into the chain to get the product $\lambda_{\ell_1,\psi} \lambda_{\ell_2,\psi}$. Note that the circles in the fusion trees wrap non-trivial cycles.}
\label{FA}
\end{figure}

As can be seen in the finite length quantum mechanical chain, the operators $Y_{\ell}$ do not generally form a group; instead, they form a fusion algebra. To understand this statement, we nucleate a pair of anyonic lines, $\ell_{1,2}$, inside of the chain as shown in Fig \ref{FA}. We can then evaluate the action of this pair on the states in two different ways.

One way is to first fuse $\ell_1$ into the chain and then do the same with $\ell_2$ thus giving us an operator $Y_{\ell_2} \circ Y_{\ell_1}$. Another way is to first fuse $\ell_2$ with $\ell_1$ and then fuse the result into the chain. Since $\ell_1$ and $\ell_2$ form a fusion algebra
\be
\ell_1 \times \ell_2 = \sum_{\ell_3} N^{\ell_3}_{\ell_1\ell_2} \ell_3~,
\ee
then $Y_{\ell_1}$ and $Y_{\ell_2}$ form a representation of this algebra on the Hilbert space of the anyonic chain
\be\la{ops}
Y_{\ell_1} \circ Y_{\ell_2} = \sum_{\ell_3} N_{\ell_1\ell_2}^{\ell_3} Y_{\ell_3}~.
\ee
Therefore, as emphasized in \cite{Aasen:2016dop}, the $Y_{\ell}$ are not required to form a group. To make this statement more explicit, we note that in general the $Y_{\ell}$ need not be invertible as matrices (i.e., they may, as in the case of the $Y_{\sigma}$ topological symmetry of the Ising model, have zero eigenvalues). Moreover, even if such inverses do exist, they are not necessarily part of the fusion algebra. Indeed, if $Y_{\ell}^{-1}$ does not represent a simple element of $\CC_{\rm in}$ or an element that can be written as a positive semi-definite integer linear combination of the representations of simple elements, then $Y_{\ell}^{-1}$ does not exist in the fusion algebra. In most of the examples we will consider, this situation will occur for at least one $\ell\in \mathcal C_{\rm in}$.

\section{The topological symmetry / topological defect correspondence}\label{correspondence}

In this section, we would like to discuss the transition from the quantum mechanical picture presented above to the CFT description discussed in the sections below. As discussed in \cite{pfeifer2012translation,Aasen:2016dop}, the topological symmetries give crucial constraints on this transition.

To understand why the topological symmetries play such an important role, note that the algebra in \eqref{ops} holds independently of the length of the chain, $L$ (as long as the length $L$ chain exists). In particular, this algebra is a property of both the finite length chain and of the theory in the thermodynamic limit. Moreover, the set of allowed topological quantum numbers is determined {\it entirely} by $\CA_{\rm in}$ and is {\it independent} of the details of the Hamiltonian.

Indeed, we need only assume that the $Y_{\ell}$ are some operators on the Hilbert space of the chain and that there exists a suitable Hamiltonian commuting with the $Y_{\ell}$. Let $|\psi\rangle$ be any eigenstate of this Hamiltonian. Applying the operator equation \eqref{ops} to $\psi$ we get an overconstrained system of polynomial equations\footnote{A similar approach has been used independently in \cite{Aasen2}.}
\be\la{lambda}
\lambda_{\ell_1,\psi} \lambda_{\ell_2,\psi} = \sum_{\ell_3} N^{\ell_3}_{\ell_1\ell_2} \lambda_{\ell_3,\psi}~,
\ee
where $\lambda_{\ell,\psi}$ is the eigenvalue of $Y_{\ell}$ on the state $|\psi\rangle$. In general, this system does not have solutions since there are more equations than unknowns. However, due to the properties of fusion tensor categories, the system \eqref{lambda} has several solutions. Since we have not specified anything about the state $|\psi\rangle$ there will be as many solutions as there are topological sectors (i.e., as there are consistent sets of topological quantum numbers). If the input fusion category, $\CC_{\rm in}$, is modular (meaning the corresponding $S$-matrix, $S^{\rm in}$, is non-degenerate and can be taken to be unitary), then it is a simple exercise in the application of Verlinde's formula \cite{Verlinde:1988sn,Pasquier:1987xj} to see that the $\lambda_{\ell,m}$ are given (up to the action of an automorphism) by
\be\label{verlinde}
\lambda_{\ell,m}={S^{\rm in}_{\ell m^*}\over S^{\rm in}_{0m^*}}~.
\ee
This expression coincides with (225) of \cite{kitaev2006anyons} (here $m^*$ is the representation of the fusion category that is conjugate to $m$, i.e., the representation whose fusion with $m$ contains the identity).

\begin{figure}
\begin{center}
\begin{tikzpicture}[pertDefect/.style={circle,draw=blue!100,thick,inner sep=0pt,minimum size=20mm,snake it},defectL/.style={circle,draw=blue!100,thick,inner sep=0pt,minimum size=30mm},HOL/.style={circle,draw=blue!100,thick,inner sep=0pt,minimum size=21.5mm,dashed},HOS/.style={circle,draw=blue!100,thick,inner sep=0pt,minimum size=15mm},auto]
\tikzset{snake it/.style={decorate, decoration=snake}}
\draw \boundellipse{0,0}{3}{1};
\draw \boundellipse{0,0}{2}{.667};
\draw[black] (0,1) -- (0,4);
\node[] at (0,4.3) {$\tilde\ell$};
\draw[black] (0,3.5) -- (.7,3.8);
\node[] at (.7,4.1) {$\tilde\ell$};
\node[] at (-.5,3.2) {$\tilde x_{L-1}$};
\draw[black] (0,2.9) -- (.7,3.2);
\node[] at (.7,3.5) {$\tilde\ell$};
\node[] at (-.3,2.4) {$\vdots$};
\draw[black] (0,1.8) -- (.7,2.1);
\node[] at (.7,2.4) {$\tilde\ell$};
\node[] at (-.3,1.5) {$\tilde x_{1}$};
\node[] at (-2.5,1) {$\tilde x_{0}$};
\node[] at (-2.3,0) {$\ell$};
\end{tikzpicture}
\caption{The number of flux sectors through the loop is invariant under the $F$ moves that take us from the configuration in Fig. \ref{TopSym} to the present one. The circle labeled by $\tilde x_0$ represents a topologically non-trivial cycle.}
\label{Ftopsymm}
\end{center}
\end{figure}
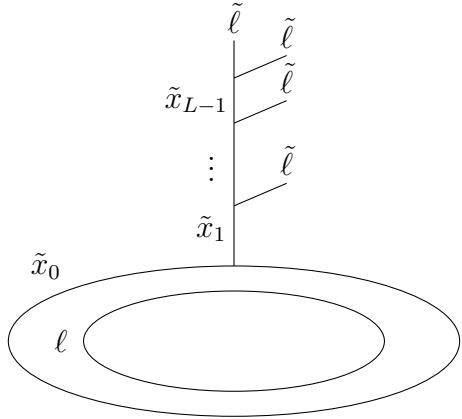

Note that our discussion does not imply that the $Y_{\ell}$ are necessarily independent operators since there could be non-trivial relations between them (we will discuss such an example in section \ref{su2infinity}). However, if the input theory is modular, then the $Y_{\ell}$ will be independent if all topological sectors are present in the theory. One quick way to count the number of topological sectors in this case is to perform a sequence of $F$-moves that transform the diagram in Fig. \ref{TopSym} into the one in Fig. \ref{Ftopsymm}. If $\tilde x_0$ takes on all values in $\CC_{\rm in}$, then we expect the topological symmetries to be independent, since the number of different flux sectors through the loop doesn't change after applying an $F$-transformation. This argument does not require eigenvectors of $Y_{\ell}$ to have definite $\tilde x_0$.\footnote{To make the previous paragraph more rigorous, we can study the \lq\lq inside" and \lq\lq outside" bases in section IIB.5 of \cite{pfeifer2012translation} and note that we work in the \lq\lq inside" basis. On the other hand, the identical diagram to Fig. \ref{Ftopsymm} written in the \lq\lq outside" basis gives a one-to-one relation between the values of the resulting $\tilde x_0^{\rm out}$ and topological flux sectors via a relation of the form
\begin{equation}
\lambda_{\ell, \tilde x_0^{\rm out}}={S^{\rm in}_{\ell\tilde x_0^{\rm out*}}\over S^{\rm in}_{0\tilde x_0^{\rm out*}}}~.
\end{equation}
Therefore, since the fusion rules do not change when we go from \lq\lq inside" to \lq\lq outside," the \lq\lq outside" basis gives a proof of our claim that if $\tilde x_0$ takes on any value in $\CC_{\rm in}$ and $\CC_{\rm in}$ is modular, we have all topological sectors present and the $Y_{\ell}$ are independent.}

A sufficient condition for this situation to occur is if there exists an integer $N$ such that fusion product $\tilde\ell^N$ contains the identity (we will see this condition is satisfied in the $\CC_{\rm in }={\rm Rep}\left[su(2)_k\right]$ examples below). Indeed, in this case, we find an infinite number of diagrams with $\tilde x_1=0$ and so $\tilde x_0$ can take on any value in $\CC_{\rm in}$, since $N^0_{\tilde x_0\tilde{x}_0^*}=1$. Another sufficient condition is if $\CC_{\rm in}$ is simple (in the sense that it doesn't contain any subsets of elements that are closed under fusion) so that $\tilde x_1$ (and therefore $\tilde x_0$) can take on any value in $\CC_{\rm in}$ (alternatively, we can assume that, as in the subset of the $\CC_{\rm in }={\rm Rep}\left[su(2)_k\right]$ examples we study below, $\tilde\ell$ is not part of a closed sub-algebra of $\CA_{\rm in}$).

Therefore, in what follows, we will assume 
\begin{itemize}
\item Our systems have independent $Y_{\ell}$.\footnote{One reason for caution if there are relations between the $Y_{\ell}$ it that, in general, the remaining $Y_{\ell}$ may satisfy fusion rules with non-integer coefficients. In this case, we are not sure if the map $Y_{\ell}\to D_{\ell}$ will lead to well-defined defect Hilbert spaces. Another subtlety is that the algebra the $Y_{\ell}$ satisfy (modulo relations) need not be equivalent to $\CA_{\rm in}$. We hope to study such theories in greater detail soon.}
\item The input fusion category, $\CC_{\rm in}$, is modular.\footnote{Our argument above uses modularity. In the non-modular case, we would need a different argument than the one given above to understand when the $Y_{\ell}$ are independent. For example, even though the non-modular $\CC_{\rm in}={\rm Rep}[su(2)_4^{\rm int}]$ chain has  ${S_{2i}\over S_{0i}}=1$ for all the simple elements labeled by the spins $i=0,1,2$, it is not true that $Y_2=Y_0$.}
\item The $Y_{\ell}$ do not map us between different Hilbert spaces\footnote{We thank D.~Aasen for a discussion of this point.} (examples of such situations include the topological symmetries corresponding to half-integer spins in the ${\rm Rep[su(2)_{4}]}$ spin-1 chains of \cite{gils2013anyonic}).
\end{itemize}

Given this discussion, we would like to understand which objects the $Y_{\ell}$ map to in the thermodynamic limit. We will assume that this limit can be described in terms of the variables of some CFT (possibly with various relevant terms turned on)---i.e., that we are in conformal perturbation theory.\footnote{We will also assume that the CFT has a unique vacuum sector.} In particular, we get the following map
\begin{equation}\label{thermoH}
H\to L_0+\bar L_0-{c\over12}+\lambda^I\oint\Phi_I~,
\end{equation}
where the $\lambda^I$ are possible relevant deformations of the critical theory.

Since the $Y_{\ell}$ are operators and not states in the quantum mechanical chain, we do not expect them to correspond to local operators in the thermodynamic limit. Instead, they should correspond to non-local operators, $D_{\ell}$. Moreover, since the $Y_{\ell}$ commute with the Hamiltonian and rotations of the spatial circle, we expect that, at the critical point
\be
\left[D_{\ell},L_0 + \bar L_0 \right]=\left[D_{\ell},L_0-\bar L_0\right]=0\quad \Rightarrow \quad \left[D_{\ell},L_0  \right]=\left[D_{\ell},\bar L_0\right]=0\,.
\ee
Furthermore, as we have argued above, the fact that the $Y_{\ell}$ satisfy the $\CA_{\rm in}$ fusion algebra for any $L$ for which the chain is well-defined tells us that the corresponding defects also satisfy the $\CA_{\rm in}$ fusion algebra
\begin{equation}\label{fusioninout}
Y_{\ell_1} \circ Y_{\ell_2}=\sum_{\ell_3}N^{\ell_3}_{\ell_1\ell_2}Y_{\ell_3}\ \ \ \Rightarrow\ \ \ D_{\ell_1}D_{\ell_2}=\sum_{\ell_3}N^{\ell_3}_{\ell_1\ell_2}D_{\ell_3}~.
\end{equation}
The fact that these operator multiplications are well-defined and independent of position strongly suggests that the $D_{\ell}$ are in fact topological defects \cite{Petkova:2000ip}---i.e., operators that can be freely deformed in correlation functions as long as they don't cross local operator insertions (besides insertions of operators in the identity module) or other defect lines. Moreover, this argument suggests that the $D_{\ell}$ form a representation of the closed sub-algebra $\CA'\simeq\CA_{\rm in}$ discussed in the introduction.

From a physical perspective, our main question is to understand when the microscopic symmetries of the chain force the relevant perturbations in \eqref{thermoH} to vanish (i.e., when we must have $\lambda^I=0$). Since the quantum mechanical model has the topological symmetries $Y_{\ell}$, the allowed perturbations in \eqref{thermoH} must be invariant under the corresponding $D_{\ell}$. In other words 
\be\la{comm}
\left[D_{\ell}, \Phi_I\right] = 0~, \ \ \ \forall \ell\in\CA'~.
\ee
The above equations turn out to be quite subtle for the most general topological defects. In the next section, we will show (Theorem \ref{Thm1}) that \eqref{comm} implies that $\Phi_I$ is in the same topological sector as the identity operator, i.e.
\be\la{QN}
\lambda_{\ell, I} = \lambda_{\ell,0}~, \ \ \ \forall \ell\in \CA'~,
\ee
where the $\lambda_{\ell,0}$ are the topological quantum numbers of the ground state (the ground state corresponds to the conformal vacuum and, therefore, to the identity operator).

Previous works used the criterion \eqref{QN} to rule out certain perturbations \cite{feiguin2007interacting} of the CFT that emerge in the thermodynamic limit (and to argue that the chain is stable to perturbations). 
In the next section, we will prove that \eqref{QN} implies \eqref{comm} for topological defects implementing discrete symmetries (like the $\mathbb{Z}_2$ spin-reflection symmetry in the Ising model). In the case of more general topological defects, \eqref{QN} is only a necessary condition for \eqref{comm} to hold. In the sequel we will formulate the conditions (arising from Theorem \ref{Thm2}, unitarity in the form of Theorem \ref{Thm3}, and the quantum mechanical constraints discussed in the introduction) under which \eqref{QN} is also a sufficient condition. This logic has a loophole: for some non-unitary chains (or even for some unitary chains coming from non-modular $\CC_{\rm in}$), it might be possible to find theories for which $\left[D_{\ell},\Phi_I\right]\ne0$ even though $\lambda_{\ell,I}=\lambda_{\ell,0}$ $\forall\ell\in\CA'$.

Finally, let us note that, in addition to the topological symmetries, it is interesting to consider the translational symmetry of the chain. It is often the case that translational symmetry corresponds to a discrete symmetry of the critical theory. Since discrete symmetries can also be generated by topological defects, our approach gives a uniform way to understand the consequences of microscopic symmetries on the effective theory (see \cite{Aasen:2016dop, Aasen2} for further details on the link between translational symmetry and discrete symmetries of the critical theory).

\section{Topological defects and the effective CFT description}
In this section, we explore some of the consequences that follow from the identification of topological symmetries with topological defects in the $L\to\infty$ limit of the anyonic chain (see also \cite{Aasen:2016dop,Aasen2} for further consequences and applications). In the first subsection, we prove that the commutator of a defect with a bulk field is non-vanishing if the bulk field has defect eigenvalue different from that of the identity, and, in the process, we give a CFT argument for half the topological protection mechanism discussed in \cite{feiguin2007interacting} (throughout we will assume that the output CFT, $\CT_{\rm out}$, has a single identity sector, i.e., $\CT_{\rm out}$ is not a direct product of different CFTs).

The converse of this theorem is much more non-trivial. Indeed, we first formulate a partial converse that depends not just on the defect eigenvalue of the bulk operator in question but also on the spectrum of defect fields. We then mention an example that shows this second condition need not be implied by the first. Furthermore, we are not aware of a general CFT proof showing that the defect commutator with a bulk field is determined by the field's defect eigenvalue (and we leave the task of proving this statement or finding a counterexample as an open problem).

However, we argue that if $\CT_{\rm out}$ is unitary, then, for any defects descending from topological symmetries of an anyonic chain, the vanishing of the defect commutator with a bulk field is determined by the topological sector of the bulk field. In order to prove this latter statement, we first argue that, under relatively mild assumptions, topological symmetries are mapped to defects that preserve the full chiral algebra of $\CT_{\rm out}$ (to demonstrate this claim, we will show that certain twist and defect fields of the CFT can be understood directly in the anyonic chain).

We then conclude by proving a theorem that constrains the possible modular input fusion algebras giving rise to a CFT via the anyonic chain mechanism. We leave further explorations of the space of such CFTs to future work.

\subsection{Defining the defect commutator and a theorem on the non-trivial defect eigenvalue sector}
Before explaining how to evaluate the commutators in \eqref{comm}, we should make the notion of a topological defect more precise. Following \cite{Petkova:2000ip} (see also \cite{Recknagel:2013uja}), we define a topological defect, $D_{A}$, to be an operator commuting with the full left and right Virasoro algebras of $\CT_{\rm out}$
\begin{equation}\label{Vircomm}
\left[L_n, D_{A}\right]=\left[\bar L_n, D_{A}\right]=0~,\ \ \ \forall\ n\in\mathbb{Z}~.
\end{equation}
This definition implies that the defect is \lq\lq transparent" to the CFT stress tensor:\footnote{At least for the theories we will be interested in below, these operators can also be thought of as Verlinde loop operators (see \cite{Verlinde:1988sn} for a definition and \cite{Alday:2009fs,Drukker:2010jp} for a discussion of the relation). The process of looping the anyon, $\ell$, through the ring is then the discrete version of the operation performed in \cite{Verlinde:1988sn}.} we are free to deform the tensionless $D_{A}$ inside a correlation function of local operators without changing the correlator as long as we don't cross local operator insertions (besides the stress tensor and other elements of the Virasoro vacuum character).\foot{Note that, when there is an extended chiral algebra (e.g., a super Virasoro or $W$-algebra), the defect may or may not satisfy
\begin{equation}\label{Wcomm}
\left[W_n^i, D_A\right]=\left[\bar W_n^i, D_A\right]=0~,\ \ \ \forall\ n\in\mathbb{Z},~i~,
\end{equation}
where the $W_n^i$ and $\bar W_n^i$ are the modes of this extended algebra.
}

An extension of Schur's lemma then implies that the defect takes the form
\begin{equation}\label{defectdef}
D_{A}=\sum_L\Delta_{AB}\cdot|B\rangle\langle B|~,
\end{equation}
where the sum is taken over (pairs of left and right) representations of Virasoro, and $D_A$ acts with the same eigenvalue on each element in the representation. We then see that the independent $D_A$ are in one-to-one correspondence with the (pairs of left and right) Virasoro representations present in the theory. In the next sections we will see that modular covariance of the CFT in the presence of topological defects implies further constraints on the coefficients, $\Delta_{AB}$, but for now we simply note that the $D_A$ act like projectors.

If the sum in \eqref{defectdef} is finite, then the theory is a Virasoro minimal model. More generally, if we have an RCFT with extended left and right chiral algebras (e.g., $W$-algebras), then the defects that commute with the full left and right chiral algebras satisfy \eqref{defectdef} with the sum taken over the finite number of (pairs of left and right) representations of the extended chiral algebras (and the labels in \eqref{defectdef} then refer to left and right pairs of representations of these chiral algebras). For simplicity, we will assume that $D_A$ commutes with the full left and right chiral algebras of the theory (the extension of some of our results to more general topological defects is straightforward). Moreover, as we mentioned above, under certain plausible assumptions, the defects we are most interested in---those that come from topological symmetries of an anyonic chain---will commute with the full left and right chiral algebras.

We are now in a position to study the commutator, $\left[D_A,\Phi_B(z,\bar z)\right]$. For simplicity, we will take $\Phi_B$ to be a chiral algebra primary (the extension to the case of descendants is trivial). Since we are particularly interested in the commutator of the defect with possible deformations of the Hamiltonian, we evaluate
\begin{equation}\label{fundcomm}
\left[D_A,\oint d\theta\Phi_B(z,\bar z)\right]=\oint d\theta\left[D_A,\Phi_B(z,\bar z)\right]~,
\end{equation}
where the integral is over a spatial circle in radial quantization. We evaluate the commutator by demanding that when we insert it in correlation functions, there are no local fields (or other defects) located radially between $D_A$ and $\Phi_B(z,\bar z)$. Moreover, to define the commutator, it suffices to determine its action on a complete set of states inserted at the origin, $|C_k\rangle\equiv\Phi_{C_k}(0,0)|0\rangle$ (by construction, we require that there are no additional local operators or other defects located between the commutator and the origin). Here $C_0\equiv C$ is the primary, and the $k\ne0$ label the descendants.

\begin{figure}
\begin{center}
\begin{tikzpicture}[pertDefect/.style={circle,draw=blue!100,thick,inner sep=0pt,minimum size=20mm,snake it},defect/.style={circle,draw=blue!100,thick,inner sep=0pt,minimum size=40mm},defectL/.style={circle,draw=blue!100,thick,inner sep=0pt,minimum size=40mm},defectS/.style={circle,draw=blue!100,thick,inner sep=0pt,minimum size=3mm},HOL/.style={circle,draw=blue!100,thick,inner sep=0pt,minimum size=21.5mm,dashed},HOS/.style={circle,draw=blue!100,thick,inner sep=0pt,minimum size=18.5mm,dashed},auto]
\tikzset{snake it/.style={decorate, decoration=snake}}
\node[defectL] (2) at (2.5,0) [shape=circle] {$\bullet$};
\node[HOS] (3) at (2.5,0) [shape=circle] {$\bullet$};
\node[] at (2.9,0) {$\Phi_C$};
\node[] at (5.,0) {$-$};
\node[defectS] at (6.5,0) {$\bullet$};
\node[] at (2.2,-.5) {$\oint\Phi_B$};
\node[] at (3.,1.6) {$D_{A}$};
\node[] at (6.9,0) {$\Phi_C$};
\node[] at (6.1,.3) {$D_{A}$};
\node[] at (5.2,-.5) {$\oint\Phi_B$};
\node[HOL] (3) at (6.5,0) [shape=circle] {$\bullet$};
\node[] at (8.5,0) {$=$};
\node[defect] (3) at (11.5,0) [shape=circle] {$\bullet$};
\node[] at (9.3,1.2) {$D_{A}$};
\node[] at (10.45,0) {$N_{BC}^E\left(\Phi_E\right)$};
\node[] at (14.,0) {$-$};
\node[HOS] (3) at (15.5,0) [shape=circle] {$\bullet$};
\node[] at (15.,-1.2) {$\Delta_{AC}\oint\Phi_B$};
\node[] at (15.9,0) {$\Phi_C$};
\end{tikzpicture}
\caption{Our definition of the commutator, $[D_{A},\oint\Phi_B]$, acting on the state $|C\rangle$. In the second diagram on the RHS we have acted on the field $\Phi_C$ using the defect.}
\label{fig:SimpleComm}
\end{center}
\end{figure}
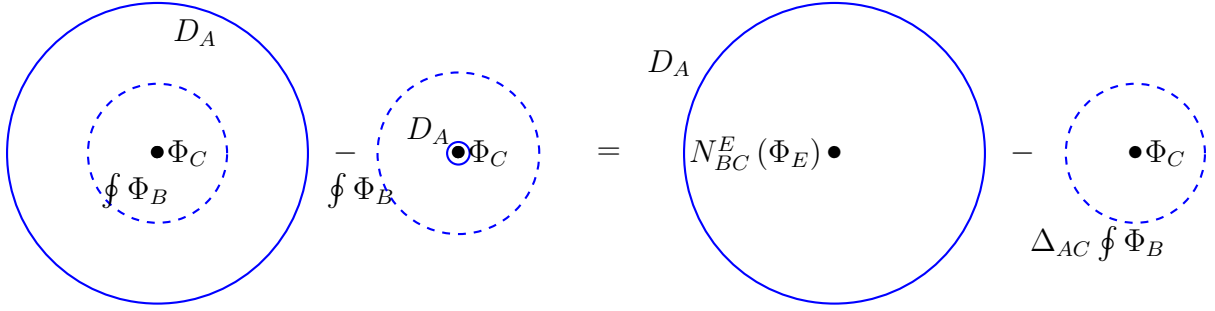

We can then evaluate the commutator as in Fig. \ref{fig:SimpleComm}: using the topological nature of the defect, we are free to shrink it to within a small distance of the origin, as in the second diagram on the LHS and then act on the state at the origin as in the second diagram on the RHS. In the first diagram on the RHS, we act with $\Phi_B$ on the state at the origin.\footnote{Note that the topological nature of the commutator allows us to evaluate \eqref{commeval} in various topologically equivalent ways (see, for example, the equivalent limit defined in \cite{Runkel:2010ym}).} Altogether, we obtain the following expression
\begin{eqnarray}\label{commeval}
\oint d\theta\left[D_A,\Phi_B(z,\bar z)\right]|C_k\rangle&=&\oint d\theta\left[D_A,\Phi_B(z,\bar z)\right]\Phi_{C_k}(0)|0\rangle\cr&=&\oint d\theta\left(\sum_{E,k'}\left(\Delta_{AE}-\Delta_{AC}\right)N_{BC}^{E}\cdot\lambda_{BC_k}^{E_{k'}}\cdot z^{\rho^{E_{k'}}_{BC_k}}\bar z^{\bar\rho^{E_{k'}}_{BC_k}}\Phi_{E_{k'}}(0)\right)|0\rangle\cr&=&{2\pi}\sum_{E,k'}\left(\Delta_{AE}-\Delta_{AC}\right)N_{BC}^{E}\cdot\lambda_{BC_k}^{E_{k'}}r^{2\rho^{E_{k'}}_{BC_k}}\Phi_{E_{k'}}|0\rangle\cr&=&{2\pi}\sum_{E,k'}\left(\Delta_{AE}-\Delta_{AC}\right)N_{BC}^{E}\cdot\lambda_{BC_k}^{E_{k'}}r^{2\rho^{E_{k'}}_{BC_k}}|E_{k'}\rangle~,
\end{eqnarray}
where the index \lq\lq $k'$" runs over the chiral algebra descendants of $\Phi_E$, $\rho^{E_{k'}}_{BC_k}=h_{E_{k'}}-h_B-h_{C_k}$, and $\bar\rho^{E_{k'}}_{BC_k}=\bar h_{E_{k'}}-\bar h_B-\bar h_{C_k}$. By definition of the fusion numbers, $N^E_{BC}$, we have that
\begin{equation}\label{nonzero}
N^E_{BC}\ne0\ \Leftrightarrow\ \exists\ k\ {\rm s.t.}\ \sum_{k'}\lambda_{BC_k}^{E_{k'}}r^{2\rho^{E_{k'}}_{BC_k}}\Phi_{E_{k'}}\equiv\left(\Phi_E\right)\ne0~,
\end{equation}
and so
\begin{equation}\label{cond}
\left[D_A,\oint d\theta\Phi_B(z,\bar z)\right]\ne0\ \Leftrightarrow\ \exists \ C,E \ {\rm s.t.} \ \left(\Delta_{AE}-\Delta_{AC}\right)N_{BC}^{E}\ne0~.
\end{equation}

Given this groundwork, we can prove the following elementary theorem for CFT operators that have defect eigenvalues different from those of the identity

\smallskip
\noindent
{\bf Theorem \ref{Thm1}:} \label{Thm1} Consider an operator, $\Phi_B$, in an RCFT, $\CT_{\rm out}$, and consider the set of topological defects, $D_A$, of $\CT_{\rm out}$. If $\exists\  A$ such that $\Delta_{AB}\ne\Delta_{A0}$ (where \lq\lq $0$" refers to the vacuum), then it follows that $\left[D_A,\oint\Phi_B\right]\ne0$.

\smallskip
\noindent
{\bf Proof:} Let us prove the contrapositive: suppose that $\left[D_A,\oint\Phi_B\right]=0$ $\forall\ A$, and let us show that $\Delta_{AB}=\Delta_{A0}$ $\forall\ A$. To that end, we see from \eqref{cond} that the vanishing of the commutator implies that
\begin{equation}\label{conseqThm1}
\left(\Delta_{AE}-\Delta_{AC}\right)N_{BC}^{E}=0~, \ \ \ \forall\ C,E~.
\end{equation}
Take $C=0$. Then, by the rules of fusion categories, $N_{B0}^E=\delta_B^E$. Therefore, we have that $\Delta_{AB}=\Delta_{A0}$ as desired.

\smallskip
\noindent
{\bf q.e.d.}
\smallskip

As we will see below, the converse of this theorem is non-trivial and requires additional conditions for general defects because of the existence of certain fields localized on these defects. However, for group-like defects (i.e., those defects whose product law is that of a discrete group and whose actions on local operators implement internal symmetries of the CFT), the converse is always true since the $\Delta_{AB}$ are genuine charges of the theory and $\Delta_{AB}=\Delta_{A0}=1$ means that $\Phi_B$ is neutral under the corresponding symmetry. At the level of defect fields, we will argue this statement follows from the fact that group-like defects have at most one such field with the quantum numbers of $\Phi_B$.

Let us conclude this subsection by noting the implications for our anyonic chain discussion: we see that if $D_{\ell}$ is a topological defect descending from an ancestor topological symmetry, $Y_{\ell}$, and if $\Phi_B$ comes from some state in the chain that has $Y_{\ell}$ eigenvalue different from that of the identity, then the deformation
\begin{equation}\label{Hdef}
\delta H_{\rm eff} \sim \lambda^B\oint d\theta\Phi_B~,
\end{equation}
is forbidden in the effective theory (similar results hold if the deformation is charged under a discrete symmetry of the chain). The reason is that the relation $Y_{\ell}\to D_{\ell}$ implies that
\begin{equation}\label{evaleq}
\lambda_{\ell, B} = \Delta_{\ell B}~,
\end{equation}
and therefore Theorem \ref{Thm1} rules out \eqref{Hdef} by the state-operator map in the CFT.

As we have mentioned above, we would like to understand the converse of this statement: suppose that $\Phi_B$ comes from a quantum mechanical state that is in the topologically trivial sector of all the $Y_{\ell}$ acting on the chain (and is uncharged under any discrete symmetries). Then, we would like to know whether the deformation in \eqref{Hdef} is ruled out or not.

\subsec{A theorem on fields in the trivial defect eigenvalue sector}
In this section, we will prove a partial converse for Theorem \ref{Thm1} and explore its consequences for anyonic chains. However, in order to set the stage for this theorem, we will first need to introduce the constraints that modular covariance of $\CT_{\rm out}$ in the presence of topological defects places on the $D_A$ \cite{Petkova:2000ip}.

To that end, consider the torus partition function of $\CT_{\rm out}$
\begin{equation}\label{TPfn}
Z=\sum_{I,J} Z_{I\bar J}\chi_I(\tilde q)\bar\chi_{\bar J}(\tilde q)~.
\end{equation}
This partition function is a trace over the Hilbert space
\begin{equation}
\CH=\oplus_{I, \bar J}Z_{I\bar J}R_{I}\otimes\bar R_{\bar J}~,
\end{equation}
where $R_I$ and $\bar R_{\bar J}$ are representations of the left and right chiral algebras.\footnote{Note that the $I, \bar J$ may in general run over multiple copies of a given representation.}

We will argue below that the RCFTs we are most interested in are those that are diagonal with respect to some maximal chiral algebra, i.e., those theories that have 
\begin{equation}
Z_{I\bar J}=\delta_{I^*\bar J}~,
\end{equation}
where $I^*$ denotes the representation conjugate to $I$, or those theories that are related to diagonal ones by an automorphism of the theory that preserves the form of defects commuting with the full chiral algebra. We will call this case the \lq\lq Cardy case" or say that such RCFTs are \lq\lq Cardy-like."

To compute the partition function in the presence of a series of defects wrapping the spatial direction, we treat the defects as operators and insert them into \eqref{TPfn}
\begin{equation}\label{TPfn2}
Z_{A_1,\cdots, A_n}=\sum_{I}\left(\prod_{\alpha}\Delta_{A_{\alpha}I}\right)\chi_I(\tilde q)\bar\chi_{I^*}(\tilde q)~.
\end{equation}
On the other hand, we can also imagine that the defects wrap the time direction and hence modify the Hilbert space of the theory. We can obtain such a configuration by applying a modular $S$-transformation to \eqref{TPfn}
\begin{equation}\label{STPfn}
Z_{A_1,\cdots, A_n}\to Z_{A_1,\cdots, A_n}'=\sum_{I,K,L} \left(\prod_{\alpha}\Delta_{A_{\alpha}I}\right)S_{IK}\bar S_{I^*L}\chi_K(\tilde q)\bar\chi_{L}(\tilde q)~,
\end{equation}
where $S_{IK}$ is the modular $S$ matrix. The consistency condition of \cite{Petkova:2000ip} requires that the transformed partition function in \eqref{STPfn} has a suitable Hilbert space interpretation, i.e., that the coefficients of the partition function are positive semi-definite integers. This condition constrains the allowed eigenvalues, $\Delta_{A_{\alpha}I}$. Note that the Hilbert space in the presence of the defects counts the fields that can live on these defects.

\subsubsec{Fields living on defects}
In this subsection, we discuss the different fields that can live on defects by considering the transformations in \eqref{STPfn} for differing numbers of defect insertions. In the next subsection, we consider twist fields (i.e., fields living at the end of defects), while in the following subsection we study defect fields (i.e., fields living in the interior of defects), and in the subsequent subsection, we consider junction fields (in our case, those that sit at the intersection of three defects).\footnote{Note that it is possible to consider more general junction fields that sit at the intersection of more than three defects. However, we will not need such configurations in our analysis below.}

We will soon argue that the defects descending from topological symmetries preserve the full left and right chiral algebras of $\CT_{\rm out}$. Therefore, in what follows, we will restrict our attention to defects commuting with the full left and right chiral algebras. The resulting defects have eigenvalues of the form
\begin{equation}\label{Deval}
\Delta_{AI} = {S_{AI^*}\over S_{0I^*}}={\bar S_{AI}\over \bar S_{0I}}={\bar S_{AI}\over S_{0I}}~,
\end{equation}
where $\bar S$ denotes complex conjugation and we have used standard properties of modular $S$-matrices (see, e.g., chapter 10 of \cite{DiFrancesco:1997nk}) in unitary theories (unitarity allows us to conclude that $S_{0I}>0$). In particular, we see that for defects descending from topological symmetries, \eqref{Deval} has the same index structure as the expression in \eqref{verlinde} and (225) of \cite{kitaev2006anyons} (this similarity is to be expected since we have a map from input objects to objects satisfying the output fusion sub-algebra $\CA_{\rm in}\simeq\CA'\subset\CA$, and our arguments around \eqref{fusioninout} suggested that the eigenvalues of $Y_{\ell}$ and $D_{\ell}$ should coincide). 

Given this discussion, it is straightforward to check that \eqref{Deval} implies that the corresponding defects satisfy the usual Verlinde fusion algebra of $\CT_{\rm out}$
\begin{equation}\label{Dverlinde}
D_{I}D_{J}=\sum_KN^K_{IJ}D_K~,
\end{equation}
where $N_{IJ}^K$ are the full fusion coefficients of $\CT_{\rm out}$.\footnote{More general topological defects satisfy a generalized algebra called the defect classifying algebra \cite{Fuchs:2010hk}.} In particular, we see that for the subset of defects coming from the topological symmetries, we have an expression of the form \eqref{fusioninout} and that these defects form a closed subalgebra.

\bigskip
\centerline{\it Twist Fields}
\bigskip

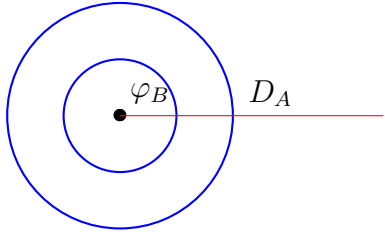
\begin{figure}
\begin{center}
\begin{tikzpicture}[pertDefect/.style={circle,draw=blue!100,thick,inner sep=0pt,minimum size=20mm,snake it},defectL/.style={circle,draw=blue!100,thick,inner sep=0pt,minimum size=30mm},HOL/.style={circle,draw=blue!100,thick,inner sep=0pt,minimum size=21.5mm,dashed},HOS/.style={circle,draw=blue!100,thick,inner sep=0pt,minimum size=15mm},auto]
\tikzset{snake it/.style={decorate, decoration=snake}}
\node[defectL] (2) at (2.5,0) [shape=circle] {$\bullet$};
\node[HOS] (3) at (2.5,0) [shape=circle] {$\bullet$};
\node[] at (2.9,.3) {$\varphi_B$};
\node[] at (4.5,.3) {$D_A$};
\draw[red] (2.5,0) -- (6,0);
\end{tikzpicture}
\caption{The twist field, $\varphi_B$, sits at the end of defect, $D_A$, at $0\in\mathbb{C}^1$. The annulus is the projection of a torus onto the plane. Therefore, to compute the Hilbert space of twist fields, we should insert $D_A$ in the torus partition sum and perform an $S$-transformation.}
\label{fig:Twist}
\end{center}
\end{figure}

A twist field, $\varphi_B$, sits at the end of a defect, $D_A$. As mentioned above and described further in Fig. \ref{fig:Twist}, we can compute the Hilbert space of twist fields by inserting $D_A$ in the partition function and performing an $S$-transformation
\begin{equation}\label{twistHilbert}
Z_{A}=\sum_I\Delta_{AI}\cdot\chi_I(\tilde q)\bar\chi_{I^*}(\tilde q)~ \ \ \ \rightarrow\ \ \ Z_{A}'=\sum_{I,K,L}\Delta_{AI}S_{IK}\bar S_{I^*L}\cdot\chi_K(q)\bar \chi_{L}(q)~.
\end{equation}
In order for the twist fields to have a suitable Hilbert space interpretation, it must be the case that \cite{Petkova:2000ip}
\begin{equation}\label{PZ}
\sum_I\Delta_{AI}S_{IK}S_{IL}\in\mathbb{Z}_{\ge0}~,  \ \ \ \forall\ K,L~.
\end{equation}
In the case of our defects of interest, we have according to \eqref{Deval} that $\Delta_{AI}=\bar S_{AI}/S_{0I}$, and so Verlinde's formula implies that the multiplicity of defect fields is given by
\begin{equation}
\sum_I\Delta_{AI}S_{IK} S_{IL}=N_{KL}^A\ge0~.
\end{equation}
In particular, the Hilbert space of twist fields on $D_A$ is
\begin{equation}\label{twistH}
\CH_{\rm twist}^A=\left(R_K\otimes \bar R_{L}\right)^{\oplus N_{KL}^{A}}~,
\end{equation}
where the multiplicity is just the fusion number, $N_{KL}^A$.

\bigskip
\centerline{\it Defect Fields}
\bigskip

\begin{figure}
\begin{center}
\begin{tikzpicture}[pertDefect/.style={circle,draw=blue!100,thick,inner sep=0pt,minimum size=20mm,snake it},defectL/.style={circle,draw=blue!100,thick,inner sep=0pt,minimum size=30mm},HOL/.style={circle,draw=blue!100,thick,inner sep=0pt,minimum size=21.5mm,dashed},HOS/.style={circle,draw=blue!100,thick,inner sep=0pt,minimum size=15mm},auto]
\tikzset{snake it/.style={decorate, decoration=snake}}
\node[defectL] (2) at (2.5,0) [shape=circle] {$\bullet$};
\node[HOS] (3) at (2.5,0) [shape=circle] {$\bullet$};
\node[] at (2.9,.3) {$\varphi_B$};
\node[] at (4.5,.3) {$D_A$};
\node[] at (.4,.3) {$D_{A^*}$};
\draw[red] (-1,0) -- (6,0);
\end{tikzpicture}
\caption{The defect field, $\varphi_B$, sits inside a defect, $D_A$, at $0\in\mathbb{C}^1$. The annulus is the projection of a torus onto the plane. Therefore, to compute the Hilbert space of defect fields, we should insert $D_A$ and its conjugate, $D_A^{\dagger} = D_{A^*}$ in the torus partition sum and perform an $S$-transformation.}
\label{fig:Defect}
\end{center}
\end{figure}
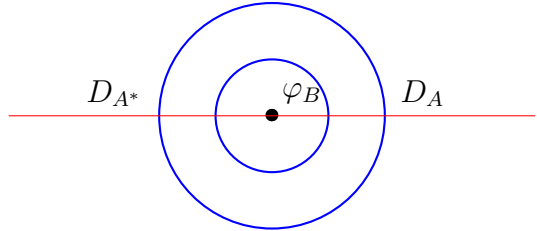

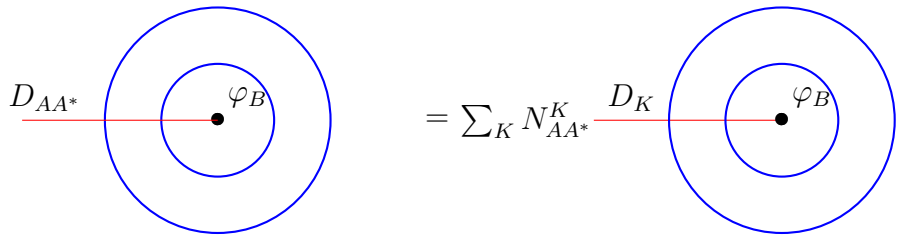
\begin{figure}
\begin{center}
\begin{tikzpicture}[pertDefect/.style={circle,draw=blue!100,thick,inner sep=0pt,minimum size=20mm,snake it},defectL/.style={circle,draw=blue!100,thick,inner sep=0pt,minimum size=30mm},HOL/.style={circle,draw=blue!100,thick,inner sep=0pt,minimum size=21.5mm,dashed},HOS/.style={circle,draw=blue!100,thick,inner sep=0pt,minimum size=15mm},auto]
\tikzset{snake it/.style={decorate, decoration=snake}}
\node[defectL] (2) at (2.5,0) [shape=circle] {$\bullet$};
\node[HOS] (3) at (2.5,0) [shape=circle] {$\bullet$};
\node[] at (2.9,.3) {$\varphi_B$};
\node[] at (10.4,.3) {$\varphi_B$};
\node[] at (.2,.3) {$D_{AA^*}$};
\draw[red] (-.1,0) -- (2.5,0);
\node[] at (5.5,0) {$=\ $};
\node[] at (6.6,0) {$\sum_KN_{AA^*}^K$};
\node[] at (8,.3) {$D_{K}$};
\draw[red] (7.5,0) -- (10,0);
\node[defectL] (2) at (10,0) [shape=circle] {$\bullet$};
\node[HOS] (3) at (10,0) [shape=circle] {$\bullet$};
\end{tikzpicture}
\caption{The spectrum of defect fields, $\varphi_B$, can alternatively be gotten by deforming the RHS of the defect in Fig \ref{fig:Defect} and bringing it to overlap with the LHS. This produces the defect $D_{AA^*}$.}
\label{fig:Defecttotwist}
\end{center}
\end{figure}

A defect field, $\varphi_B$, sits in the interior of a defect, $D_A$. As described in Fig. \ref{fig:Defect}, to compute the space of such fields, we should insert $D_A$ and $D_{A^*}$ into the partition function and perform an $S$ transformation (we can also consider inserting two non-conjugate defects if we want to compute the spectrum of defect fields that change the defect from one type to another)
\begin{equation}\label{DefectHilbert}
Z_{A,A^*}=\sum_I\Delta_{AI}\Delta_{A^*I}\cdot\chi_I(\tilde q)\bar\chi_{I^*}(\tilde q)~ \ \ \ \rightarrow\ \ \ Z_{A,A^*}'=\sum_{I,K,L}\Delta_{AI}\Delta_{A^*I}S_{IK}\bar S_{I^*L}\cdot\chi_K(q)\bar \chi_{L}(q)~.
\end{equation}
According to the manipulations in Fig. \ref{fig:Defecttotwist}, the Hilbert space of defect fields is given by
\begin{equation}\label{defectH}
\CH_{\rm defect}^A=\left(R_P\otimes \bar R_Q\right)^{\oplus\sum_K N_{AA^*}^{K}N_{PQ}^K}~.
\end{equation}

\bigskip
\centerline{\it Junction Fields}
\bigskip

\begin{figure}
\begin{center}
\begin{tikzpicture}[pertDefect/.style={circle,draw=blue!100,thick,inner sep=0pt,minimum size=20mm,snake it},defectL/.style={circle,draw=blue!100,thick,inner sep=0pt,minimum size=30mm},HOL/.style={circle,draw=blue!100,thick,inner sep=0pt,minimum size=21.5mm,dashed},HOS/.style={circle,draw=blue!100,thick,inner sep=0pt,minimum size=15mm},auto]
\tikzset{snake it/.style={decorate, decoration=snake}}
\node[defectL] (2) at (2.5,0) [shape=circle] {$\bullet$};
\node[HOS] (3) at (2.5,0) [shape=circle] {$\bullet$};
\node[] at (2.9,.3) {$\varphi_B$};
\node[] at (4.5,.3) {$D_C$};
\node[] at (-.2,.3) {$D_{A^*}$};
\node[] at (3,2.) {$D_E$};
\draw[red] (2.5,0) -- (3.5,2);
\draw[red] (-1,0) -- (6,0);
\end{tikzpicture}
\caption{The junction field, $\varphi_B$, sits at the intersection of three defects, $D_{A^*,C,E}$, at the point $0\in\mathbb{C}^1$. The annulus is the projection of a torus onto the plane. Therefore, to compute the Hilbert space of junction fields, we should insert $D_{A^*}$, $D_C$, and $D_E$ in the torus partition sum and perform an $S$-transformation. Note that by setting these various defects to the trivial defect, we can recover any of the other types of fields we discussed before.}
\label{fig:Junction}
\end{center}
\end{figure}
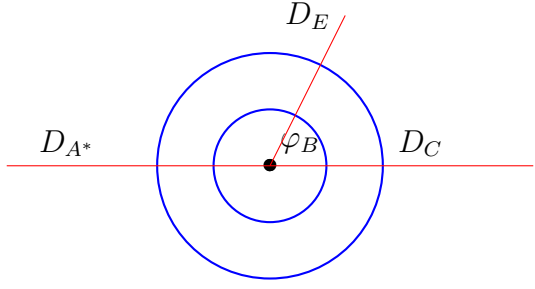

A junction field, $\varphi_B$, sits at the intersection of three defects, $D_{A^*,C,E}$ (by choosing some of these defects to be the identity, we get defect fields and twist fields). According to the discussion in Fig. \ref{fig:Junction}, we compute the resulting Hilbert space of junction fields by inserting $D_{A^*}D_{C}D_E$ into the torus partition sum
\begin{equation}\label{DefectHilbert}
Z_{A^*,C,E}=\sum_I\Delta_{A^*I}\Delta_{CI}\Delta_{EI}\cdot\chi_I(\tilde q)\bar\chi_{I^*}(\tilde q)~ \ \ \ \rightarrow\ \ \ Z_{A^*,C,E}'=\sum_{I,K,L}\Delta_{A^*I}\Delta_{CI}\Delta_{EI}S_{IK}\bar S_{I^*L}\cdot\chi_K(q)\bar \chi_{L}(q)~.
\end{equation}
We can find the defect Hilbert space by performing two iterations of the move described in Fig. \ref{fig:Defecttotwist}. Doing so, we obtain
\begin{equation}\label{junctionH}
\CH_{\rm junction}^{A^*,C,E}=\left(R_P\otimes\bar R_Q\right)^{\sum_{K,F}N_{A^*C}^KN_{KE}^FN_{PQ}^F}~.
\end{equation}

\subsec{Precursors of defect fields in the anyonic chain and symmetries preserved by the $D_{\ell}$}\label{interlude}
The results in the previous subsection apply to the Cardy case (i.e., the case in which $\CT_{\rm out}$ is diagonal with respect to some maximal chiral algebra or can be written as a diagonal chiral algebra up to an automorphism that preserves the form of the topological defects commuting with the full chiral algebra). In this section, we will argue that the output CFT from the anyonic chain construction should be of this form and that the $D_{\ell}$ defects should commute with the full output chiral algebra. Our argument follows from identifying and counting quantum mechanical ancestors of some of the fields discussed in the previous subsection. In particular, we will specialize to fields and defects descending directly from input fusion objects (related results have been obtained in \cite{Aasen:2016dop, Aasen2}).\footnote{These are the fields and defects realizing $\CA'$.}

\begin{figure}
\centerline{\includegraphics[width=\linewidth]{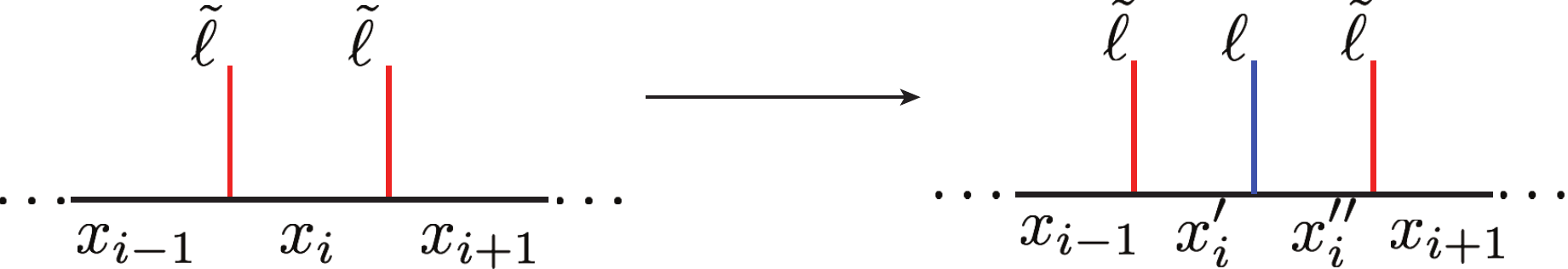}}
\caption{ On the LHS, we have a segment of the anyonic chain. To derive the chain ancestor of the twist fields sitting at the end of $D_{\ell}$, we insert a corresponding anyon, $\ell$, orthogonally to the chain on the RHS (the authors of the upcoming \cite{Aasen2} independently consider similar insertions). The anyonic chain ancestor of the twist field transforming in representation $(x_i',x_i''^*)$ is in one-to-one correspondence with the degrees of freedom living on the two links of the chain intersecting $\ell$.}
\label{fig:twistchain}
\end{figure}

To that end, let us first construct the chain ancestor of the twist fields living at the end of a defect, $D_{\ell}$. We would like to insert this defect along the time direction and consider its precursor in the quantum mechanics. The only anyonic chain object we can insert in the time direction that has the quantum numbers of $D_{\ell}$ is the anyon, $\ell$. We can locally attach $\ell$ to the chain (see Fig. \ref{fig:twistchain}) in $N_{x_i'x_i''^*}^{\ell}$ different ways.\footnote{Formally, these $N_{x_i'x_i''^*}^{\ell}$  attachments need not all be globally consistent for finite $L$. However, we will assume that $\tilde\ell$ and $\CC_{\rm in}$ are chosen such that these attachments are globally consistent for finite $L$. For example, we can choose $\CC_{\rm in}$ to be simple or demand that $\tilde\ell$ is not part of a closed fusion sub-algebra of $\CA_{\rm in}$. It would be interesting to further study models in which not all of the attachments in Fig. \ref{fig:twistchain} are consistent for finite $L$. We thank D.~Aasen for a discussion of this point.} In fact, according to our expression in \eqref{twistH}, this is precisely the number of twist fields in representation $(x_i',x_i''^*)$. Therefore, we see that the degrees of freedom localized on the part of the chain that intersects $\ell$ map to twist fields of the CFT.

\begin{figure}
\centerline{\includegraphics[width=\linewidth]{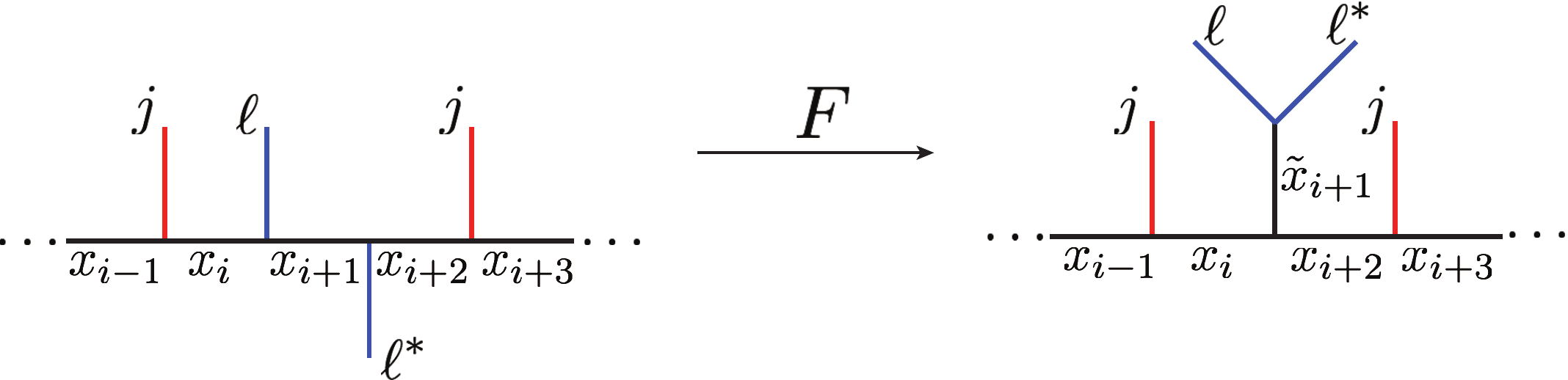}}
\caption{On the LHS, we have inserted conjugate anyons. This figure is the quantum mechanical analog of Fig. \ref{fig:Defect}. In particular, the links labeled by $x_i$ and $x_{i+2}$ are in one-to-one correspondence with the quantum mechanical ancestors of the defect fields of type $(x_i, x_{i+2}^*)$. To get to the RHS, we perform an $F$ transformation (we neglect additional non-vanishing pre-factors that accompany the $F$ transformation since we are only interested in counting the number of defect fields of type $(x_i, x_{i+2}^*)$).}
\label{fig:defectchain}
\end{figure}

We can now repeat this logic with defect fields descending from input objects. As described in Fig. \ref{fig:defectchain}, we insert conjugate anyons $\ell$ and $\ell^*$. The chain analogs of the defect fields are then in one-to-one correspondence with the links labeled $x_i$ and $x_{i+2}$ that bound the two defects. After performing an $F$ transformation, it is easy to see that the multiplicity of ways we can attach the two defects to the chain is just
\begin{equation}\label{chaindefectmult}
\sum_{\tilde x_{i+1}}N_{\ell\ell^*}^{\tilde x_{i+1}}N_{x_ix_{i+2}^*}^{\tilde x_{i+1}}\,.
\end{equation}
This expression coincides precisely with the spectrum of defect fields of type $(x_{i},x_{i+2}^*)$ in \eqref{defectH}.

Note that there were two assumptions that went into the computations leading to \eqref{twistH} and \eqref{defectH}: the defects in question commute with the full chiral algebra and $\CT_{\rm out}$ is of Cardy type. Since we find matching numbers of degrees of freedom in the chain itself (at least for those degrees of freedom that are directly related to $\CC_{\rm in}$), we have found non-trivial evidence for the claim that the output CFT and the $D_{\ell}$ defects really have these properties. Moreover, since modularity of the CFT played a crucial role in deriving \eqref{twistH} and \eqref{defectH}, we see that in a rather precise sense, the anyonic chain has knowledge of this modular characteristic of the output theory.

\subsec{A partial converse of Theorem \ref{Thm1}}
Given the above construction of fields localized to defects, we are now in a position to prove the following partial converse to Theorem \ref{Thm1}

\smallskip
\noindent
{\bf Theorem \ref{Thm2}:} \label{Thm2} Consider a local bulk operator, $\Phi_B$, in a \lq\lq Cardy-like" RCFT, $\CT_{\rm out}$, and consider a topological defect, $D_A$, commuting with the maximal left and right chiral algebras of $\CT_{\rm out}$. If {\bf(i)} $\Delta_{AB}=\Delta_{A0}$ and {\bf(ii)} there is at most one field with the same quantum numbers (under the maximal left and right chiral algebras) as $\Phi_B$ living on $D_A$, then $\left[D_A,\oint\Phi_B\right]=0$.

The conditions imposed on $\CT_{\rm out}$ and $D_A$ are motivated by the discussion in section \ref{interlude} and the fact that we are particularly interested in the case in which $\CT_{\rm out}$ is the output RCFT of some anyonic chain with $D_A=D_{\ell}$ descending from a topological symmetry, $Y_{\ell}$. However, our proof of Theorem \ref{Thm2} does not assume an anyonic chain origin of the RCFT.

\begin{figure}
\centerline{\includegraphics[width=\linewidth]{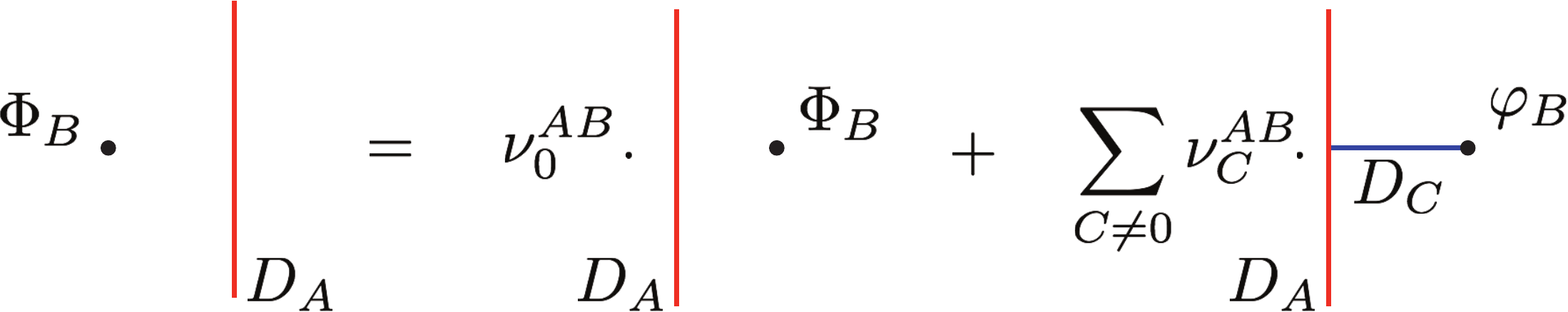}}
\caption{We sweep the defect $D_A$ across the local operator, $\Phi_B$. On the RHS, we obtain a linear combination of a bulk field and a set of twist fields, $\varphi_B$.}
\label{fig:Sweep}
\end{figure}

To prove Theorem \ref{Thm2}, we take a different approach to evaluating the commutator, $\left[D_A, \Phi_B\right]$ (we can equivalently use the methods in \cite{Frohlich:2006ch}). To that end, consider passing the defect, $D_A$, across the field, $\Phi_B$, as in Fig. \ref{fig:Sweep}. When the bulk field $\Phi_B$ is swept by $D_A$, it emerges as a linear combination (with complex coefficients, $\nu_C^{AB}$) of a bulk field on the other side (the first term on the RHS of Fig. \ref{fig:Sweep}) and a series of twist fields coupling to topological flux tubes, $D_C$.

All terms on the RHS of Fig. \ref{fig:Sweep} must have certain features. First, the terms are independent of other operator insertions sufficiently far away (i.e., Fig. \ref{fig:Sweep} is an operator equation). Second, the asymptotics of the LHS must be unchanged in the transition: at infinity we should only have the defect, $D_A$, and the local operators on the RHS should have the same quantum numbers as $\Phi_B$. Finally, the defect junction on the RHS should be topological. In particular, this means that the junction field sitting at the junction of the $A$ and $C$ type defects should have the quantum numbers of the identity (i.e., there cannot be an additional local coordinate dependence).

With these observations, we are ready to compute the set of possible $D_C$ flux tubes and twist fields appearing in Fig. \ref{fig:Sweep}. First of all, according to \eqref{twistH}, the number of $\varphi_B$ twist fields for a defect of type $D_C$ is $N_{BB^*}^C$. Next, the allowed set of $D_C$ that connect topologically to $D_A$ is obtained by taking \eqref{junctionH}, setting $P=Q=0$, and taking $C\to A$, $E\to C$. This set of fields is just
\begin{equation}\label{numtube}
\sum_{K,F}N^K_{A^*A}N^F_{KC}N_{00}^F=\sum_KN^K_{A^*A}N^0_{KC}=N^{C^*}_{A^*A}=N^C_{AA^*}~.
\end{equation}
As a result, the number of terms, $|C|$, on the RHS of Fig. 7 (including the identity) is
\begin{equation}
|C|=\sum_{C}N^C_{AA^*}N^C_{BB^*}=\left|\CH_{\rm defect}^{A, R_B\otimes\bar R_{B^*}}\right|~,
\end{equation}
where, in the last equality, we have noted that this is just the number of defect fields of type $\varphi_B$ sitting on the defect $D_A$. Therefore, we learn that the possible fields appearing on the RHS of Fig. \ref{fig:Sweep} are in one-to-one correspondence with the corresponding defect fields on $D_A$. In particular, we see that
\begin{equation}
|C|=\left|\CH_{\rm defect}^{A, R_B\otimes\bar R_{B^*}}\right|\ge1~.
\end{equation}

\begin{figure}
\centerline{\includegraphics[width=\linewidth]{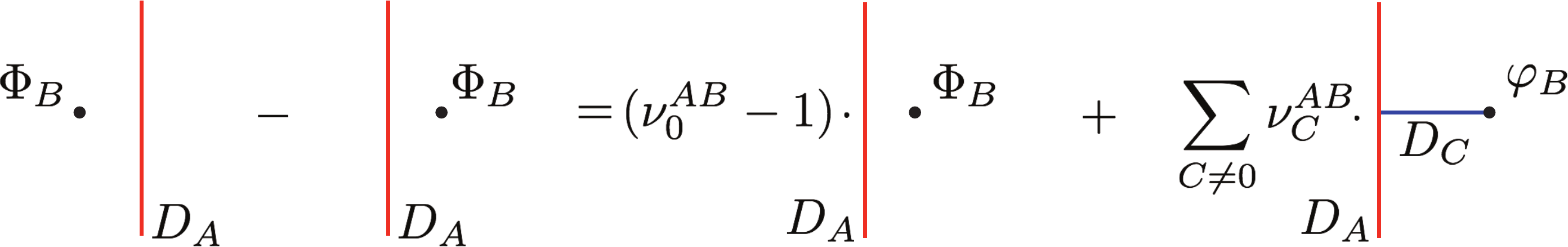}}
\caption{We use the diagrams in Fig. \ref{fig:Sweep} to compute the defect commutator, $[D_A, \Phi_B]$.}
\label{fig:SweepC}
\end{figure}

Now, let us consider the commutator $\left[D_A, \Phi_B\right]$ in Fig. \ref{fig:SweepC}. Clearly, if $\nu_0^{AB}=1$ and $|C|=\left|\CH_{\rm defect}^{A, R_B\otimes\bar R_{B^*}}\right|=1$, then all the diagrams on the RHS of Fig. \ref{fig:SweepC} vanish. Indeed, taking $|C|=\left|\CH_{\rm defect}^{A, R_B\otimes\bar R_{B^*}}\right|=1$ means that there are no $C\ne0$ contributions on the RHS. As a result, $\left[D_A, \Phi_B\right]=0$.

More generally, suppose $|C|=1$ but $\nu_0^{AB}$ is not necessarily equal to one. Clearly, we can evaluate $\nu_0^{AB}$ by closing the defect on the LHS of Fig. \ref{fig:Sweep} into a circle that doesn't surround $\Phi_B$ and doing the same on the RHS except now enclosing $\Phi_B$. We find
\begin{equation}\label{result}
\nu_0^{AB}={\left({S_{A0}\over S_{00}}\right)\over\left({S_{AB^*}\over S_{0B^*}}\right)}~.
\end{equation}
In particular, we see that $\nu^{AB}_0=1$ is equivalent to the statement that the defect eigenvalue of $B$ is the same as the defect eigenvalue of the identity. Altogether, we have proven our theorem
\begin{equation}\label{theorempf}
|C|=\left|\CH_{\rm defect}^{A, R_B\otimes\bar R_{B^*}}\right|=1~, \ \ \ {S_{A0}\over S_{00}}={S_{AB^*}\over S_{0B^*}}~\ \ \ \Rightarrow\ \ \ \ \left[D_A, \Phi_B\right]=0~.
\end{equation}
{\bf q.e.d.}

\bigskip
\noindent
Let us now make some comments on Theorem \ref{Thm2}. If the defect, $D_A$, is group-like, then $N_{AA^*}^C=\delta_0^C$ (this statement is due to the fact that $D_A$ and $D_{A^*}$ implement the actions of inverse group elements). As a result, $\left|\CH_{\rm defect}^{A, R_B\otimes \bar R_{B^*}}\right|=1$, and, as mentioned in the paragraph below \eqref{conseqThm1}, the theorem simplifies: $S_{AB^*}/S_{0B^*}=S_{A0}/S_{00}$ implies that $\left[D_A, \Phi_B\right]=0$. Note the interpretation of this theorem is simple. The commutator of an operator with a defect vanishes if there are no bound states that can be generated on the defect that mix with the operator and if the local operator cannot acquire a multiplicative factor when passing through the defect. This discussion connects with a topologically equivalent way of evaluating the commutator due to Runkel that also involves defect fields \cite{Runkel:2010ym}. Indeed, we can see explicitly how the existence of defect fields affects the commutator.

At the level of modular tensor categories (MTCs), the quantum double of the discrete group $S_3$, $D(S_3)$, provides an example of a theory in which condition {\bf(i)} of Theorem \ref{Thm2} is satisfied but condition {\bf(ii)} is not. In the notation of \cite{beigi2011quantum}, we can identify $C\leftrightarrow\Phi_B$ and $D_A\leftrightarrow D_C$ (alternatively, we can identify $F\leftrightarrow\Phi_B$ and $D_A\leftrightarrow D_F$). Indeed, we have that $S_{CC^*}/S_{0C^*}=S_{C0}/S_{00}$, but $\sum_KN_{CC^*}^KN_{CC^*}^K=3$. Note, however, that $[D_C,C]=0$. This commutator vanishes because the $\nu^{AB}_K$ defined in Fig. \ref{fig:Sweep} vanish. It would be interesting to consider quantum doubles of more exotic categories to see if one can find an example in which the commutator is also non-vanishing.

Finally, let us discuss this theorem from the perspective of the anyonic chains we introduced above. Suppose that we have a quantum mechanical state, $|\Phi_I\rangle$, in the anyonic chain that has $Y_{\ell}$ eigenvalues equal to those of the identity for all $\ell$ (and is uncharged under any discrete symmetries), i.e., it is in the topological sector of the identity. We see that Theorem \ref{Thm2} gives us hope that even if $\Phi_I$ is a relevant operator in $\CT_{\rm out}$, it may still be possible for $[D_{\ell}, \Phi_I]\ne0$ if condition {\bf(ii)} of Theorem \ref{Thm2} is not satisfied. This result would imply that we can rule out the deformation $\delta H_{\rm eff}\sim\lambda\oint\Phi_I$ even though $|\Phi_I\rangle$ is in the topologically trivial sector.

We are not aware of a proof implying these defect commutators must vanish for general topological defects in $\CT_{\rm out}$. However,  we will see in the next section that unitarity combined with modularity and the quantum mechanical conditions satisfied by the chains rule this possibility out for the descendants of topological symmetries and fields in the trivial topological sector.

\subsec{A fusion theorem}
In order to better understand the role the quantum mechanics of the anyonic chain plays in constraining the defect commutator discussed in \eqref{comm}, we will need to supplement Theorem \ref{Thm2} with additional results on fields transforming in the trivial defect eigenvalue sector. In particular, it will be useful to prove the following statement

\smallskip
\noindent
{\bf Theorem \ref{Thm3}:} \label{Thm3} Consider a unitary CFT, $\CT_{\rm out}$. Suppose $\Phi_I$ and $\Phi_J$ satisfy $S_{AI^*}/S_{0I^*}=S_{A0}/S_{00}$ and $S_{AJ^*}/S_{0J^*}=S_{A0}/S_{00}$ respectively. Then, any $\Phi_K$ appearing in the fusion of $\Phi_I$ and $\Phi_J$ satisfies $S_{AK^*}/S_{0K^*}=S_{A0}/S_{00}$.\footnote{We thank T. Gannon for discussions related to this proof. Similar statements have appeared in the literature on subfactor theory \cite{liu2016exchange,jiang2016noncommutative}.}

\smallskip
\noindent
{\bf Proof:} We have
\begin{equation}\label{initobs}
{S_{AI^*}\over S_{0I^*}}{S_{AJ^*}\over S_{0J^*}}=\left({S_{0A^*}\over S_{0I^*}}{S_{0A^*}\over S_{0J^*}}\right)\left({S_{IA^*}\over S_{0A^*}}{S_{JA^*}\over S_{0A^*}}\right)=\left({S_{0A}\over S_{0I}}{S_{0A}\over S_{0J}}\right)\sum_K N_{IJ}^K{S_{KA^*}\over S_{0A^*}}~.
\end{equation}
Taking the absolute value, using the fact that the first row (column) in the modular $S$-matrix is positive for a unitary theory, and using the triangle inequality, we find
\begin{equation}
\left|{S_{AI^*}\over S_{0I^*}}{S_{AJ^*}\over S_{0J^*}}\right|=\left|\left({S_{0A}\over S_{0I}}{S_{0A}\over S_{0J}}\right)\sum_K N_{IJ}^K{S_{KA^*}\over S_{0A^*}}\right|\le\left({S_{0A}\over S_{0I}}{S_{0A}\over S_{0J}}\right)\sum_K N_{IJ}^K\left|{S_{KA^*}\over S_{0A^*}}\right|~.
\end{equation}
The LHS of the above equation has the following special property that follows from the conditions of the theorem and the positivity (all we actually need here is the weaker condition of reality) of $S_{A0}/S_{00}$ in a unitary theory
\begin{equation}\label{absval}
\left|{S_{AI^*}\over S_{0I^*}}{S_{AJ^*}\over S_{0J^*}}\right|=\left|{S_{A0}\over S_{00}}{S_{A0}\over S_{00}}\right|={S_{A0}\over S_{00}}{S_{A0}\over S_{00}}={S_{AI^*}\over S_{0I^*}}{S_{AJ^*}\over S_{0J^*}}~.
\end{equation}
Moreover, we have that
\begin{equation}
{S_{A0}\over S_{00}}{S_{A0}\over S_{00}}=\left({S_{0A}\over S_{0I}}{S_{0A}\over S_{0J}}\right)\left({S_{0I}\over S_{00}}{S_{0J}\over S_{00}}\right)=\left({S_{0A}\over S_{0I}}{S_{0A}\over S_{0J}}\right)\sum_KN_{IJ}^K{S_{K0}\over S_{00}}~.
\end{equation}
Therefore, we see
\begin{equation}
\sum_KN_{IJ}^K{S_{K0}\over S_{00}}\le\sum_K N_{IJ}^K\left|{S_{KA^*}\over S_{0A^*}}\right|~.
\end{equation}
On the other hand, since $S_{KA^*}/S_{0A^*}$ are eigenvalues of fusion matrices (which have positive semi-definite coefficients), we have that, in a unitary theory, $|S_{KA^*}/S_{0A^*}|\le S_{K0}/S_{00}$ and so in fact
\begin{equation}
\sum_KN_{IJ}^K{S_{K0}\over S_{00}}\le\sum_K N_{IJ}^K\left|{S_{KA^*}\over S_{0A^*}}\right|\le\sum_KN_{IJ}^K{S_{K0}\over S_{00}}~.
\end{equation}
As a result, we conclude that
\begin{equation}
\sum_K N_{IJ}^K{S_{KA^*}\over S_{0A^*}}=\sum_KN_{IJ}^K{S_{K0}\over S_{00}}=\sum_K N_{IJ}^K\left|{S_{KA^*}\over S_{0A^*}}\right|~,
\end{equation}
and so
\begin{equation}
{S_{AK^*}\over S_{0K^*}}={S_{A0}\over S_{00}}~.
\end{equation}
{\bf q.e.d.}

We can now use Theorem \ref{Thm2} to prove a constraint on the $\CT_{\rm out}$ modular $S$-matrix elements corresponding to objects that descend from $\CC_{\rm in}$. In the next subsection, we will use this constraint to prove the vanishing of \eqref{comm}.

To that end, we assume that the $Y_{\ell}$ satisfy the criteria discussed below \eqref{verlinde} and are therefore independent operators. Let us now also suppose that $\Phi_I$ (in the topological sector of the identity) does not satisfy condition {\bf(ii)} of Theorem \ref{Thm2} for at least one $D_{\ell}$ representing an element of the output fusion sub algebra, $\CA'\simeq\CA_{\rm in}$, that is isomorphic to the input fusion subalgebra (here we are implicitly using our argument given below \eqref{chaindefectmult} that the $D_{\ell}$ commute with the full chiral algebra and that $\CT_{\rm out}$ is Cardy-like). In this case, we must have that the fusion $\Phi_I\Phi_I^*$ contains a non-identity element of $\CA'$, i.e. $F_I \equiv \left\{\Phi'|\Phi'\subset\Phi_I\Phi_I^*\right\}$ satisfies
\begin{equation}
\nu_I=F_I \cap\CA'\ne\left\{0\right\}~.
\end{equation}
In fact, we have the stronger statement that
\begin{equation}
\nu_{I\ell}=F_I\cap F_{\ell}\ne\left\{0\right\}~.
\end{equation}
In other words, there is some $\tilde\Phi\in\nu_I, \nu_{I\ell}$ that is non-trivial (i.e., $\tilde\Phi\ne0$). By Theorem \ref{Thm3}, this $\tilde\Phi$ satisfies
\begin{equation}\label{Phi0cond}
{S_{m\tilde\Phi^*}\over S_{0\tilde\Phi^*}}={S_{m0}\over S_{00}}~, \ \ \ \forall\ \Phi_{m}\in\CA'~.
\end{equation}
As a result, we see that the sub-block of the modular $S$ matrix of the output theory that corresponds to the objects in the input fusion category is necessarily degenerate.

As a final aside, note that Theorem \ref{Thm3} uses unitarity. Therefore, the above conclusions may not hold for some non-unitary theories. Also, they may not hold in theories in which there are relations between the $Y_{\ell}$ (unless we can guarantee that the corresponding relations lead to sensible defect Hilbert spaces upon identifying topological symmetries with topological defects).

\subsec{Further implications for anyonic chains}
Given the above groundwork, we can now prove the vanishing of \eqref{comm} for $\Phi_I$ in the topological sector of the identity.

\smallskip
\noindent
{\bf Theorem \ref{Thm4}:} \label{Thm4} Consider a unitary RCFT, $\CT_{\rm out}$, arising from some input fusion category via the thermodynamic limit of an anyonic chain (satisfying the conditions discussed below \eqref{verlinde}). For the topological defects, $D_{\ell}$, descending from topological symmetries, $Y_{\ell}$, of the anyonic chain, any local operator, $\Phi_I$, with all $D_{\ell}$ eigenvalues equal to those of the identity, satisfies $\left[D_{\ell}, \Phi_I\right]=0$ $\forall\ell$.

\smallskip
\noindent
{\bf Proof:} Let us prove by contradiction. To that end, suppose $\exists\ \ell$ such that $\left[D_{\ell}, \Phi_I\right]\ne0$. In this case, the contrapositive of Theorem \ref{Thm2} implies that $\Phi_I$ does not satisfy condition {\bf(ii)}.

Then, it must follow from \eqref{Phi0cond} that there is an operator, $\tilde\Phi\in\CA'\simeq\CA_{\rm in}$ with $\tilde\Phi\ne0$, appearing in the fusion product $\Phi_I\Phi_I^*$ satisfying (note again that we are using the argument given below \eqref{chaindefectmult} that the $D_{\ell}$ commute with the full chiral algebra)
\begin{equation}\label{Phi0cond2}
{S_{\tilde\Phi m^*}\over S_{0m^*}}={S_{0\tilde\Phi}\over S_{00}}~, \ \ \ \forall\ \Phi_{m}\in\CA'~.
\end{equation}
This equation implies that the corresponding topological symmetry, $Y_{\tilde\Phi}$, acts as a multiple of the identity on the input data
\begin{equation}
Y_{\tilde\Phi}={\bf 1}\cdot {S_{0\tilde\Phi}\over S_{00}}~.
\end{equation}
Since the remaining states in the theory that do not descend directly from the input must still be in topological sectors labeled by the input data, this discussion implies that $Y_{\tilde\Phi}$ is proportional to the identity when acting on all states, which is a contradiction since $Y_{\tilde\Phi}$ would not be an independent operator (here we use the conditions below \eqref{verlinde}).

\noindent
{\bf q.e.d.}

Our above discussion also proves the following corollary

\noindent
{\bf Corollary \ref{Thm5}:} \label{Thm5} Consider a unitary CFT, $\CT_{\rm out}$, arising from some modular input fusion category via the thermodynamic limit of an anyonic chain (satisfying the conditions discussed below \eqref{verlinde}). Let us call $\CC_{\rm out}$ the MTC associated with $\CT_{\rm out}$. If every proper fusion sub-category, $\CC'\subset\CC_{\rm out}$, has a corresponding sub-block of the modular $S$-matrix with two rows proportional to the identity row, then the input fusion category must (up to choices of Frobenius-Schur indicators\footnote{This additional data does not affect the fusion rules, but it can affect the $F$-symbols \cite{kitaev2006anyons}.}) be isomorphic to $\CC_{\rm out}$.

\section{Examples and applications}\label{Exapp}
In this section we will give some brief examples and applications of the general discussion\footnote{Let us also note that it is possible to use unitarity, aspects of the discussion in section 4, and certain topological manipulations to derive the vanishing of $\left[D_{\ell},\Phi_I\right]$ from a more topological point of view. Elements of this discussion were summarized in Appendix B of the previous version of this paper, but this appendix has evolved into a separate publication \cite{Toappear}.} above.

 \subsection{Basic example: Fibonacci ``golden'' chain}

The Fibonacci chain is the first non-trivial example of an anyonic chain \cite{feiguin2007interacting}. The input fusion category, $\CC_{\rm in}$, has an algebra consisting of two elements, $\mathcal A_{\rm in} = \mathcal A_{\rm Fib} =  \{1,\tau\}$. The fusion rules are
\be
1\times 1 =1~,\ \ \  1\times \tau = \tau~,\ \ \ \tau \times \tau = 1 + \tau~.
\ee
One simplification in this theory is that there is only one higher-dimensional $F$-symbol 
\be
F^{\tau\tau\tau}_\tau =  \begin{pmatrix}
    \phi^{-1} & \phi^{-\frac{1}{2}}  \\
    \phi^{-\frac{1}{2}} &-\phi^{-1} 
  \end{pmatrix}~,
\ee
where $\phi =\frac{1+\sqrt{5}}{2}$ is the golden ratio and is one of the solutions to the following equation
\be
\phi - \frac{1}{\phi} = 1~.
\ee
The only non-trivial choice of external particle involves taking $\tilde\ell=\tau$. In this case, the local projector $P_i^{(\tau)(1)}$ discussed around \eqref{Hanyon} has the form
\be
P_i^{(\tau)(1)} = -\begin{pmatrix}
    \phi^{-2} & \phi^{-\frac{3}{2}}  \\
    \phi^{-\frac{3}{2}} &-\phi^{-1} 
  \end{pmatrix}~,
\ee
 when acting on $|\tau1\tau\rangle$ and $|\tau\tau\tau\rangle$. The Hamiltonian is then given by
 \be\la{HFib}
 H = J\sum_i P_i^{(\tau)(1)}~,
 \ee
which energetically favors one of the fusion outcomes over the other. Such a situation is physical, because, if they exist in nature, anyons will interact and those interactions will likely lift some of the degeneracies we expect in the limit of well-separated particles.

This model is related to the fusion algebra of $su(2)_3$, and the (modular) input fusion category is sometimes referred to as $\CC_{\rm in}={\rm Rep}\left[su(2)_3^{\rm int}\right]={\rm Rep}\left[so(3)_3\right]$ (i.e., the integer spin representations of $su(2)_3$).\footnote{Note that the $so(3)_3$ WZW model is not modular \cite{Gepner:1986wi}. However, one can associate the $\CC_{\rm in}$ fusion category with the representations of the $\left(G_2\right)_1$ WZW theory (although we should emphasize that the input is not a full-fledged RCFT, but only its associated UMTC).} Let us explain this relation further since we will need it in the next set of examples we discuss. The fusion category ${\rm Rep}\left[su(2)_3\right]$ consists of four objects $\mathcal A_{su(2)_3} = \{0,\frac{1}{2}, 1, \frac{3}{2}\}$. The fusion rules are given by \cite{Gepner:1986wi}
\begin{center}
\begin{tabular}{ |c|cccc| } 
 \hline
$\times$ & 0&$\frac{1}{2}$ &$1$ &$\frac{3}{2}$ \\ 
 \hline
$0$ &0& $\frac{1}{2}$ &1 &$\frac{3}{2}$ \\ 
$\frac{1}{2}$ &$\frac{1}{2}$ & $0+1$ & $\frac{1}{2} + \frac{3}{2}$ &$1$ \\ 
 $1$ & $1$ & $\frac{1}{2} + \frac{3}{2}$& $0+1$ &$\frac{1}{2}$\\ 
 $\frac{3}{2}$ & $\frac{3}{2}$ & $1$ & $\frac{1}{2}$ &$0$\\ 
 \hline
\end{tabular}
\end{center}
Note, that, as in the case of the usual $su(2)$ algebra, the integer subset is closed under fusion. Moreover, the fusion rules of the integer subset coincide with the fusion rules of Fibonacci anyons, i.e., $\CA_{su(2)_3} = \CA_{\rm fib}$. 

Let us now discuss the topological symmetries. We have one non-trivial operator, $Y_\tau$, satisfying
\be\la{defectFib}
Y_\tau \circ Y_\tau = \mathbf{1} + Y_\tau~.
\ee
$Y_\tau$ has only two distinct eigenvalues $\lambda_{\tau,0} = \phi={S^{\rm in}_{\tau0}\over S^{\rm in}_{00}}$ and $\lambda_{\tau,\tau} = - \frac{1}{\phi}={S^{\rm in}_{\tau\tau}\over S^{\rm in}_{0\tau}}$ (where $S^{\rm in}$ is the modular $S$ matrix of the Fibonacci theory). Note that $\phi$ and $-\frac{1}{\phi}$ solve the polynomial equation \eqref{defectFib}. Moreover, as in the discussion below Fig. \ref{Ftopsymm}, it is easy to check that both topological sectors are present in the theory (and so $Y_{\tau}$ is an independent operator).
 
Although this example is particularly simple, it also illustrates an important point discussed above: $Y_\tau^{-1}$ does not belong to the fusion algebra of the topological operators. Indeed, let's suppose that $Y_\tau^{-1}$ exists in the fusion algebra and then find a contradiction. To that end, consider a state, $|\psi\rangle$, in the topologically trivial sector. Under $Y_{\tau}^{-1}$, this state has quantum number $\lambda_{\tau^{-1},\psi} = \frac{1}{\phi}$. Moreover, there must exist positive semi-definite integers $p$ and $q$ such that
\be
Y^{-1}_\tau = p\cdot \mathbf 1 + q\cdot Y_\tau~.
\ee
Now, acting with this equation on the state $|\psi\rangle$ yields
\be
\frac{1}{\phi} = p + q\phi \ \ \ \Rightarrow\ \ \  p = {1-\phi^2q\over\phi}\notin\mathbb Z~,
\ee
which contradicts the assumption that $Y_{\tau}^{-1}$ is in the fusion algebra. Thus, we see that, although $Y_{\tau}$ is invertible as a matrix, it is not invertible as a topological symmetry (hence, as emphasized in \cite{Aasen:2016dop}, the word \lq\lq symmetry" is not particularly appropriate).

In the thermodynamic limit, the dynamics of the spin chain is governed by a CFT \cite{feiguin2007interacting}. In the anti-ferromagnetic case, the CFT is the superconformal $\CM(5,4)$ minimal model, while, in the ferromagnetic case, it is the 3-state Potts model (i.e., the $\mathbb Z_3$ parafermions). For the sake of brevity, we will confine our discussion to the parafermion case (the $\CM(5,4)$ case is largely similar, with $W_3$ symmetry replaced by  $\CN=1$ superconformal symmetry).

As a Virasoro theory, the $\mathbb Z_3$ parafermions have the following primaries and dimensions
\begin{eqnarray}\label{Z3specvir}
D\left(\Phi_{(1,1),(1,1)}\right)&=&0~, \ \ \ D\left(\Phi_{(1,5),(1,5)}\right)=6~,\ \ \ D\left(\Phi_{(1,5),(1,1)}\right)=3~, \ \ \ D\left(\Phi_{(1,1),(1,5)}\right)=3~,\cr D\left(\Phi_{(2,1),(2,1)}\right)&=&{4\over5}~, \ \ \ D\left(\Phi_{(2,5),(2,5)}\right)={14\over5}~,\ \ \ D\left(\Phi_{(2,5),(2,1)}\right)={9\over5}~, \ \ \ D\left(\Phi_{(2,1),(2,5)}\right)={9\over5}~,\ \ \ \ \ \ \cr D\left(\Phi_{(3,3),(3,3)}^{i=1,2}\right)&=&{2\over15}~, \ \ \ D\left(\Phi_{(4,3),(4,3)}^{i=1,2}\right)={4\over3}~.
\end{eqnarray}
On the other hand, these primaries combine into the following $W_3$ representations
\begin{eqnarray}\label{Z3specW3}
\CO_{(0,0)}&=&\Phi_{(1,1),(1,1)}\oplus\Phi_{(1,5),(1,5)}\oplus\Phi_{(1,5),(1,1)}\oplus\Phi_{(1,1),(1,5)}~,\cr\CO_{(0,-2)}&=&\Phi_{(4,3),(4,3)}^{1}~, \ \ \ \CO_{(0,2)}=\Phi_{(4,3),(4,3)}^{2}~,\cr\CO_{(2,0)}&=&\Phi_{(2,1),(2,1)}\oplus\Phi_{(2,5),(2,5)}\oplus \Phi_{(2,5),(2,1)}\oplus\Phi_{(2,1),(2,5)}~, \cr\CO_{(2,-2)}&=&\Phi_{(3,3),(3,3)}^{1}~, \ \ \ \CO_{(2,2)}=\Phi_{(3,3),(3,3)}^{2}~, \ \ \ 
\end{eqnarray}

It is straightforward to check that the only possible map consistent with the fusion in \eqref{defectFib} and modularity of the RCFT is\footnote{Any other solution consistent with \eqref{defectFib} does not lead to a well-defined defect Hilbert space. Indeed, one finds multiplicities of defect fields that are not just positive semi-definite integers.}
\begin{eqnarray}\label{Z3Dout}
Y_{\tau}\to D_{\CO_{(2,0)}}&=&\sum_{i,j}{S_{(2,0),(i,j)}\over S_{(1,1),(i,j)}}|\CO_{(i,j)}\rangle\langle\CO_{(i,j)}|\cr&=&{1+\sqrt{5}\over2}\left(|\CO_{(0,0)}\rangle\langle\CO_{(0,0)}|+|\CO_{(0,-2)}\rangle\langle\CO_{(0,-2)}|+|\CO_{(0,2)}\rangle\langle\CO_{(0,2)}|\right)\cr&+&{1-\sqrt{5}\over2}\left(|\CO_{(2,0)}\rangle\langle\CO_{(2,0)}|+|\CO_{(2,-2)}\rangle\langle\CO_{(2,-2)}|+|\CO_{(2,2)}\rangle\langle\CO_{(2,2)}|\right)~,
\end{eqnarray}
where we have used the well-known parafermion $S$-matrix (e.g., see chapter 10 of \cite{DiFrancesco:1997nk}). This result illustrates another important point discussed above: $Y_{\tau}$ maps to a defect that commutes with the full $W_3$ chiral algebra of the $\mathbb Z_3$ parafermions (i.e., not just with Virasoro\footnote{Similarly, in the $\CM(5,4)$ case, modular covariance forces the topological defect related to $Y_{\tau}$ to commute with the full $\CN=1$ super-Virasoro algebra.}). Moreover, the 3-state Potts model is diagonal with respect to this $W_3$ symmetry (although it is not diagonal with respect to Virasoro). Above, we argued this statement was true under relatively mild assumptions for RCFTs coming from anyonic chains. In addition, note that as expected from our general discussion, $D_{\CO_{(2,0)}}$ forms a closed fusion sub-algebra that is isomorphic to $\CA_{\rm in}$
\begin{equation}\label{defectsubalg}
D_{\CO_{(2,0)}}^2=D_{\CO_{(0,0)}}+D_{\CO_{(2,0)}}~.
\end{equation}

Finally, to understand why we have critical behavior in the thermodynamic limit of the chain (i.e., why the $\lambda^I$ in \eqref{thermoH} are forced to vanish), note that the only relevant deformations in the trivial sector of $D_{\CO_{(2,0)}}$ are in the representations $\CO_{(0,\pm2)}$. However, these representations are charged under the $\mathbb Z_3$ symmetry of the CFT (which in turn descends from the microscopic $\mathbb Z_3$ symmetry of the chain). This statement can also be understood by examining the corresponding group-like defects that implement the $\mathbb Z_3$ symmetry
\begin{eqnarray}
D_{\CO_{(0,-2)}}&=&|\CO_{(0,0)}\rangle\langle\CO_{(0,0)}|+|\CO_{(2,0)}\rangle\langle\CO_{(2,0)}|+e^{2\pi i\over3}\left(|\CO_{(0,-2)}\rangle\langle\CO_{(0,-2)}|+|\CO_{(2,-2)}\rangle\langle\CO_{(2,-2)}|\right)\cr&+&e^{-{2\pi i\over3}}\left(|\CO_{(0,2)}\rangle\langle\CO_{(0,2)}|+|\CO_{(2,2)}\rangle\langle\CO_{(2,2)}|\right)~,\cr D_{\CO_{(0,2)}}&=&|\CO_{(0,0)}\rangle\langle\CO_{(0,0)}|+|\CO_{(2,0)}\rangle\langle\CO_{(2,0)}|+e^{-{2\pi i\over3}}\left(|\CO_{(0,-2)}\rangle\langle\CO_{(0,-2)}|+|\CO_{(2,-2)}\rangle\langle\CO_{(2,-2)}|\right)\cr&+&e^{{2\pi i\over3}}\left(|\CO_{(0,2)}\rangle\langle\CO_{(0,2)}|+|\CO_{(2,2)}\rangle\langle\CO_{(2,2)}|\right)~.
\end{eqnarray}
In particular, we see that $\CO_{(0,\pm2)}$ do not have the same defect eigenvalues as the identity.

 \subsection{The $su(2)_k^{\rm int}$ generalization}\label{su2kexample}
The Fibonacci chain has a straightforward generalization: a chain with $\CC_{\rm in}={\rm Rep}\left[su(2)_k^{\rm int}\right]$ and $3\le k\in\mathbb Z$ (i.e., the fusion category of integer spins of $su(2)_k$). The fusion rules for $su(2)_k$ are \cite{Gepner:1986wi}
\be\label{su2kmult}
j_1 \times j_2 = \sum_{j=|j_1-j_2|}^{{\rm min}\left(j_1+j_2, \ k-j_1-j_2\right)}j~.
\ee
As in the $k=3$ case, the integer subset forms a closed sub-algebra which we will use to construct the chain. In what follows, we will study the case of odd $k$, i.e.
\begin{equation}
k=2n+1~, \ \ \ n\in\mathbb{Z}_{\ge1}~.
\end{equation}
The case of even $k$ is not modular.

To generalize the previous example in a translationally invariant manner, we replace the $\tau$ external legs with anyons of ``angular momentum'' $j = \frac{k-1}{2}$. The fusion rules are
\be
\frac{k-1}{2} \times \frac{k-1}{2} = 0 \oplus 1\,.
\ee
In the rest of this section, we summarize the main results and leave details of the computations to the Appendix.

These anyonic chains are also critical with the corresponding critical theories being the $\mathcal M(k+2, k+1)$ minimal models (with the central charge $c =1 - \frac{6}{(k+1)(k+2)}$) in the anti-ferromagnetic case and the $\mathbb Z_k$ parafermions (with the central charge $c=2 \frac{k-1}{k+2}$) in the ferromagnetic case. As in the $\mathbb Z_3$ parafermion case discussed above, it is straightforward to check that these theories have no relevant perturbations that are invariant under the defects, $D_{\ell}$, related to the integer spin topological symmetries, except for perturbations that are charged under the discrete symmetries (i.e., $\mathbb Z_2$ in the AF case and $\mathbb Z_k$ in the F case) coming from the microscopic translational symmetry (see the Appendix). It is also easy to verify that the $Y_{\ell}^{-1}$ are not in the fusion algebra and so the corresponding defects are not group-like. Moreover, one observes that these defects commute with the full chiral algebra of $\CT_{\rm out}$ and satisfy a fusion algebra isomorphic to $\CA_{\rm in}$. Finally, note that all the theories discussed here are diagonal with respect to the maximal chiral algebra (e.g., $W_n$ in the case of the $\mathbb Z_n$ parafermions).

Before finishing this section, let us note that many different generalizations of the anyonic chains are available. For example, one can consider placing spin $\frac{k-s}{2}$ anyons on the external legs, which would produce a spin-${s\over2}$ generalization of the anyonic spin-$\frac{1}{2}$ chains. This spin-$1$ generalization was considered in \cite{gils2013anyonic}, where both gapped and gapless phases were found. In the gapped phase the topological symmetry is broken spontaneously and leads to degeneracy of the ground state as well as to fractionalized anyonic edge excitations, similarly to what happens in the AKLT model. Another natural generalization is the inclusion of the Majumdar-Ghosh type of interaction, which has been studied in spin-$\frac{1}{2}$ chains \cite{trebst2008collective} (where it also produces a rich phase diagram).

\subsection{The $su(2)_{\infty}^{\rm int}$ case}\la{su2infinity}

When $k \to \infty$, the fusion rules in \eqref{su2kmult} become the standard ones for $su(2)$
\be\label{su2reg}
j_1 \times j_2 = \sum_{j=|j_1-j_2|}^{j_1+j_2}j~.
\ee
Therefore, it is tempting to imagine that we might recover the Heisenberg spin-$\frac{1}{2}$ chain in this limit. 

However, at the level of output RCFTs, this idea doesn't appear too promising. For example, we saw above that in the ferromagnetic case, the Heisenberg chain goes to a scale invariant theory with dynamical critical exponent $z=2$. On the other hand, the $k\to\infty$ limit of the $Z_k$ parafermions is a $c=2$ conformal (and Lorentz invariant) theory, $Z_{\infty}$, described in \cite{Bakas:1990ry} (this theory is also $W_{\infty}$ symmetric). Moreover, in the anti-ferromagnetic case, the Heisenberg theory is described by the $su(2)_1$ theory, but the $k\to\infty$ limit of $\CM(k+2, k+1)$ is an irrational CFT \cite{Runkel:2001ng}.

One possible resolution to these apparent discrepancies is to imagine that the two limits $k\to\infty$ and $L\to\infty$ don't commute. However, if we try to take the limit $k\to\infty$ before taking the thermodynamic limit, the situation is also puzzling. For example, the finite length $su(2)^{\rm int}_{\infty}$ chain has an infinite dimensional Hilbert space, while the Heisenberg chain has a finite dimensional one (this fact suggests we should place additional constraints on the $su(2)^{\rm int}_{\infty}$ chain if we wish to reproduce the Heisenberg chain).

Another puzzle involves the fact that in the $su(2)_1$ model, all the non-trivial defects are group-like \cite{Fuchs:2007tx}. On the other hand, we saw that none of the topological symmetries in the $su(2)_k$ chains were group-like. Therefore, one can wonder how this state of affairs might be compatible with our proposed topological defect / topological symmetry correspondence. In short, there is no contradiction because $su(2)^{\rm int}_{\infty}$ is not modular and infinitely many of the corresponding topological symmetries, $Y_{\ell\ne0}$, are not independent. Indeed, the putative modular $S$ matrix takes the following form when acting on finite spin representations
\begin{equation}\label{Sinfinity}
S_{j_1,j_2}=\lim_{k\to\infty}\left(\sqrt{2\over k+2}\sin\left[{(2j_1+1)(2j_2+1)\pi\over(k+2)}\right]\right)=0~.
\end{equation}
Moreover, the finite spin $Y_{\ell}$ have the following action on finite spin states
\begin{equation}\label{redundant}
Y_{\ell}|j_1\rangle=(2\ell+1)|j_1\rangle~.
\end{equation}
While it is certainly true that the $Y_{\ell}$ form a representation of the fusion algebra \eqref{su2reg}
\begin{equation}
Y_{\ell_1}\circ Y_{\ell_2}=\sum_{\ell=|\ell_1-\ell_2|}^{\ell_1+\ell_2}Y_{\ell}~,
\end{equation}
we also see from \eqref{redundant} that infinitely many of these operators are not independent.

\subsection{Some constraints on $\CC_{\rm in}$ from $\CC_{\rm out}$}
In this section, we would like to use Corollary \ref{Thm5} to exhibit some constraints on the input fusion category given $\CC_{\rm out}$. We leave a more detailed study of these constraints to future work and content ourselves with presenting a few examples here.

For instance, if the output theory has a UMTC that is just $D(S_3)$, i.e., the quantum double of $S_3$, then $\CC_{\rm in}=D(S_3)$ is the only possible modular input category as well (at least as long as the conditions described below \eqref{verlinde} hold). To that end, we use the notation and results of \cite{beigi2011quantum} to point out that $D(S_3)$ has three non-trivial closed sub-algebras. One sub-algebra is generated by $A$ and $B$ (a $\mathbb Z_2$ sub-algebra), the second sub-algebra is generated by $A$, $B$, and $C$, and the final sub-algebra is generated by $A$, $B$, and $F$. From the expression of the modular $S$-matrix in (6) of \cite{beigi2011quantum}, it is straightforward to see that all of the fusion sub-categories satisfy the conditions required for Corollary \ref{Thm5} to apply.

Similar comments apply to the Ising model (where, as in the discussion in footnote \ref{isingfoot}, we do not distinguish between $\CC_{\rm in}$ that differ by Frobenius-Schur indicators). Indeed, this model contains a closed fusion sub-algebra corresponding to the $\mathbb Z_2$ symmetry of the theory
\begin{equation}\label{isingz2}
1\times 1= 1~, \ \ \ \epsilon\times 1 = \epsilon~, \ \ \ \epsilon\times\epsilon=1~.
\end{equation}
From the standard expression for the modular $S$ matrix (see, e.g., (10.138) of \cite{DiFrancesco:1997nk}), it is straightforward to see that the sub-block of the $S$ matrix corresponding to the degrees of freedom in \eqref{isingz2} have the necessary degeneracy for Corollary \ref{Thm5} to apply (although we should note that the $\mathbb Z_2$ fusion category cannot give rise to the Ising CFT on simpler grounds: the $\mathbb Z_2$ fusion algebra does not have multiple fusion channels for a given product of elements).

\section{Conclusions}
In this paper, we have explored various model-independent features of the 1D anyonic chain / 2D CFT correspondence. In particular, we discussed the topological symmetry / topological defect correspondence of \cite{pfeifer2012translation, Aasen:2016dop, Aasen2} and found new features of this relation. We then used the map between topological symmetries and topological defects to study the stability of the critical theory. We hope these results will be of some use in exploring the topological properties of general FQHE systems and also for better understanding the space of 2D CFTs.

By its nature, our work leads to many questions, including
\begin{itemize}
\item Are there RCFTs with operators that satisfy condition {\bf(i)} of Theorem \ref{Thm2} but for which $\left[D_A, \oint\Phi_B\right]\ne0$?
\item For RCFTs coming from unitary (and modular) chains, we saw that condition {\bf(ii)} of Theorem \ref{Thm2} is implied by condition {\bf(i)} in the case of defects descending from topological symmetries and local operators in the topological sector of the identity. On the other hand, there are interesting non-unitary chains that may have applications to condensed matter physics. Does such a result apply to these latter chains? If not, can non-unitary chains exhibit enhanced topological protection of gaplessness?
\item What is the space of 2D CFTs we can generate from the anyonic chain procedure? Is it possible to associate an RCFT with the (extended) Haagerup subfactor (see, e.g., \cite{izumi2001structure,Evans:2010yr,Gannon:2016bch} for some important work in this direction) thus realizing an old dream of Vaughan Jones?
\item We saw that when we take $k\to\infty$ in the $\CC_{\rm in}={\rm Rep}\left[su(2)_k^{\rm int}\right]$ case, we could end up with an irrational output theory \cite{Runkel:2001ng}. However, we are not aware of the existence of an irrational output theory if $\CC_{\rm in}$ contains a finite number of simple elements. Can one prove that this is indeed the case or find a counterexample? 
\item In the case of translationally invariant spin chains, we know that the output theory can be scale invariant, but not Lorentz invariant. What about in the case of translationally invariant anyonic chains? One possibility is that the ``topological symmetry'' implies Lorentz symmetry (at least when combined with scale invariance).
\item What happens in the case when the anyonic chain is not translationally invariant? From our discussion in section \ref{interlude}, we saw that we could interpret the insertion of defects in the output theory as anyonic inhomogeneities in the quantum mechanical chain. Will such theories still be critical (if the translationally invariant parent is) or will they flow to gapped theories (see related questions in the context of flows between RCFTs in \cite{Kormos:2009sk,Gaiotto:2012np})?
\item It would likely be worthwhile to connect our discussion directly with systems exhibiting the fractional quantum Hall effect. Some topological defects have been previously studied in the FQH literature \cite{barkeshli2013theory} \cite{teo2015theory} \cite{Gromov:2016umy}, and it would be interesting to investigate potential relations between these defects and the defects discussed in the present manuscript.
\item Can non-modular chains / chains with relations among the $Y_{\ell}$ have enhanced topological protection of criticality?
\item The anyonic chain procedure inputs associativity via the $F$-symbols, but it also contains much more information. To what extent does this construction complement more traditional approaches to CFT that impose associativity like the conformal bootstrap?
\item Are there bounds on the number of \lq\lq accidental" defects (commuting with the full left and right chiral algebras) that can emerge in the thermodynamic limit as a function of the number of simple input defects? Such a result would be similar in spirit to the conjectured bounds on accidental symmetries studied in a quite different context \cite{Buican:2011ty}.
\item Are there any potential connections between our results and conjectured bounds on numbers of relevant deformations in RG flows in various dimensions \cite{Gukov:2015qea}?
\end{itemize}

We hope to return to (some of) these questions soon.

\ack{ \bigskip
We would like to thank D.~Aasen, E.~Ardonne, F.~Chousein, N.~Drukker, T.~Gannon, M.~Levin, Z.~Liu, A.~Ludwig, W.~Nabil, T.~Nishinaka, A.~O'Bannon, S.~Ramgoolam, C.~Schweigert, G.~ Tak\'acs, A.~Tomasiello, H.~Verlinde, and G.~Watts for useful discussions and correspondence. M.~B. is grateful for the hospitality and stimulating environments of the Galileo Galilei Institute during the workshop on \lq\lq Conformal Field Theories and Renormalization Group Flows in Dimensions $d>2$," of Tsinghua University during the \lq\lq Strings 2016" conference, of the Yukawa Institute for Theoretical Physics during the workshop \lq\lq Strings and Fields 2016," of Harvard University during the \lq\lq Workshop on Subfactor Theory, Quantum Field Theory, and Quantum Information," and of the Perimeter Institute, which is supported by the Government of Canada and the Province of Ontario, during the workshop on \lq\lq Exact Operator Algebras in Superconformal Field Theories," where parts of this work were completed. M.~B.'s research is partially supported by the Royal Society under the grant \lq\lq New Constraints and Phenomena in Quantum Field Theory." A.~G. is grateful to the hospitality of the Aspen Center for Physics, which is supported by National Science Foundation grant PHY-1066293 and the 2016 Boulder Summer School for Condensed
Matter and Materials Physics supported through the NSF grant DMR-13001648. A.~G.'s research is supported by the Leo Kadanoff Fellowship.
}

\newpage
\begin{appendices}\label{appA}
\section{Spin-$1/2$ $su(2)_k^{int}$ chains}
We will briefly describe how topological protection works for the critical spin-$1/2$ chains based on the $su(2)_k^{int}$ input fusion algebra with $k$ odd.

\subsection{The anti-ferromagnetic case and the $\CM(k+2,k+1)$ minimal models}
When the Hamiltonian is AF it was found \cite{gils2009collective} that the output critical theory is the $\CM(k+2,k+1)$ minimal model. We have the following defects, $D_{(1,2m+1)}$  $(0\le m\le {k-1\over2})$, arising from the microscopic description
\begin{eqnarray}\label{Dinout}
D_{(1,2m+1)}&=&\sum_{(r,s)}{S_{(1,2m+1),(r,s)}\over S_{(1,1),(r,s)}}|(r,s),(r,s)\rangle\langle(r,s),(r,s)|\cr&=&\sum_{(r,s)}{\sin\left({2m+1\over k+2}\pi s\right)\over\sin\left({1\over k+2}\pi s\right)}|(r,s),(r,s)\rangle\langle(r,s),(r,s)|~.
\end{eqnarray}
These defects satisfy the input Verlinde algebra and form a closed subalgebra under the output fusion rules. According to the Theorem \ref{Thm1}, a deformation of the form $\delta H=\lambda\oint\Phi_{(r,s),(r,s)}$ is ruled out when
\begin{equation}\label{Mppprule}
{\sin\left({2m+1\over k+2}\pi s\right)\over\sin\left({1\over k+2}\pi s\right)}\ne{\sin\left({2m+1\over k+2}\pi \right)\over\sin\left({1\over k+2}\pi\right)}~,
\end{equation}
where $r$ and $s$ are Kac labels.

Therefore, all deformations with $s\ne1$ are ruled out. This restriction leaves deformations of the Hamiltonian by $\Phi_{(r,1),(r,1)}$ unprotected by the topological symmetries and corresponding CFT defects. In fact, the vanishing of the following defect commutator
\begin{equation}\label{vanishingcomm}
\left[D_{(1,2m+1)},\Phi_{(r,1),(r,1)}\right]=0~,
\end{equation}
is a consequence of our Theorem \ref{Thm2} described above. Indeed, it is easy to check that the condition in \eqref{theorempf} is satisfied
\begin{equation}
|C| = \sum_{(p,q)} N^{(p,q)}_{(1,2m+1),(1,2m+1)}N^{(p,q)}_{(r,1),(r,1)} = 1~, \ \ \ {S_{(1,2m+1),(r,1)}\over S_{(1,2m+1),(1,1)}}={S_{(1,2m+1),(1,1)}\over S_{(1,1),(1,1)}}~.
\end{equation}

However, since the microscopic anyonic chain has a $\mathbb{Z}_2$ symmetry (related to a translation symmetry \cite{Aasen:2016dop, Aasen2}), we should also impose this symmetry in the CFT. It is not difficult to see that in the CFT, this symmetry is generated by the defect related to $\Phi_{(1,k+1)}$ since
\begin{equation}
\Phi_{(1,k+1)}\times \Phi_{(1,k+1)}=\Phi_{(1,1)}~.
\end{equation}
Indeed, we see that
\begin{eqnarray}\label{Z2defect}
D_{(1,k+1)}&=&\sum_{(r,s)}{S_{(1,k+1),(r,s)}\over S_{(1,1),(r,s)}}|(r,s),(r,s)\rangle\langle(r,s),(r,s)|\cr&=&\sum_{(r,s)}(-1)^{kr}{\sin\left[{(1+k)^2\pi s\over2+k}\right]\over\sin\left[{(1+k)\pi s\over2+k}\right]}|(r,s),(r,s)\rangle\langle(r,s),(r,s)|\cr&=&\sum_{(r,s)}(-1)^{1+kr}|(r,s),(r,s)\rangle\langle(r,s),(r,s)|~.
\end{eqnarray}
Therefore, we can apply Theorem \ref{Thm1} to rule out $\mathbb{Z}_2$ odd deformations via the defect commutator. To that end, since $k$ is odd, when $r$ is odd, the defect eigenvalue is the same as for the identity. On the other hand, when $r$ is even, then the defect eigenvalue is $-1$. The corresponding states have non-trivial $\mathbb{Z}_2$ charge.

As a result, only CFT deformations of the form $\Phi_{(2n+1,1),(2n+1,1)}$ are allowed. For $n=0$, these are deformations in the identity module of the CFT. Clearly, such deformations cannot be relevant (the dimension zero primary deformation is trivial). For $n>0$, a similar story holds since the primary has scaling dimension
\begin{equation}
2h_{(2n+1,1)}\ge2{k+3\over k+1}>2~.
\end{equation}

\subsection{The ferromagnetic case and the $Z_k$ parafermions }
In the F case the spin chain is also critical and is described by $Z_k$ parafermions. These theories can also be described by an $su(2)_k/u(1)_k$ coset.

Now, given a coset model $G_k/H_l$ the modular $S$-matrix is given by
\be\la{CosetSmatrix}
S^{G_k/H_l}_{(a,p)(b,q)} = c_0(G,H,k,l) S^{G_k}_{ab} \bar S_{pq}^{H_l}~,
\ee
where $a$ and$ b$ are labels for the AKM primaries in the $G_k$ theory, $p$ and $q$ are labels for the AKM primaries in the $H_l$ model, and $c_0(G,H,k,l)$ is an overall constant that can be fixed by the unitarity of the $S$-matrix. 

The primary fields in this case are denoted as $\CO_{2j,2m}$ where the labels $2j \in \{0,1,\ldots,k\}$ and $2m \in \mathbb Z$ subject to the identifications
\be
\CO_{2j,2m} = \CO_{2j,2m+2k} = \CO_{2j+k,2m-k}~,  \ \ \ 2m + 2j = 0 \ \  {\rm mod} \  2~.
\ee
The fusion rules for these primaries are
\be
\CO_{2j_1,2m_1} \times \CO_{2j_2,2m_2} = \sum_{j=2|j_1-j_2|\in 2\mathbb{Z}}^{{\rm min}\left(2(j_1+j_2),2k-2(j_1+j_2)\right)} \CO_{2j,2(m_1 + m_2)}~.
\ee
In more detail, the fields of interest are 
\begin{itemize}
\item The vacuum $\CO_{0,0} = \mathbf 1$
\item The parafermion currents $\psi_m = \CO_{0,2m}$, where $m \in \{1, \ldots, k-1\}$. These fields generate the $\mathbb Z_k$ symmetry. The fusion rules form an abelian group $\mathbb Z_k$
\be \la{parafusion}
\psi_l \times \psi_k = \psi_{l+k}~.
\ee
\item The primary fields $\sigma_l = \CO_{l,l}$. 
\item The neutral $su(2)_k$ fields $\varepsilon_j = \CO_{2j,0}$. These fields form the $\CA'\simeq\CA_{\rm in}=su(2)_k^{\rm int}$ subalgebra. The fusion rules of $\varepsilon_l$ produce only neutral (in the $\mathbb Z_k$ sense) fields
\be
\label{neutralfusion}
\CO_{(2j_1,0)}\times \CO_{(2j_2,0)}=\sum_{j=2|j_1-j_2|\in 2\mathbb{Z}}^{{\rm min}\left(2(j_1+j_2),2k-2(j_1+j_2)\right)}\CO_{(2j,0)}~.
\ee
\end{itemize}
The $S$-matrix is given by applying \eqref{CosetSmatrix}, which yields
\be\la{paraSmatrix}
S_{\CO_{2j_1,2m_1}, \CO_{2j_2,2m_2}} = \frac{1}{D} \cdot  \frac{\sin\left(\frac{(2j_1+1)(2j_2+1)}{k+2}\pi\right)}{\sin\left(\frac{\pi}{k+2}\right)} \cdot \exp\left(-\frac{4\pi i}{k}m_1m_2\right)~,
\ee
where $\frac{1}{D} = \frac{1}{\sqrt{3\phi+2}} = \frac{2}{\sqrt{15}} \sin\left(\frac{\pi}{5}\right)$ is the total quantum dimension. Theorem \ref{Thm1} implies that the primary field $\CO_{2j_2,2m_2}$ is problematic if it commutes with {\it all} of the $\CA'$ defects $D_{\CO_{2j_1,0}}$. Explicitly this takes form
\be
[D_{\CO_{2j_1,0}}, \CO_{2j_2,2m_2}] = 0~, \ \ \  \forall j_1~,
\ee
which leads to the condition
\be
 \frac{\sin\left(\frac{(2j_1+1)(2j_2+1)}{k+2}\pi\right)}{\sin\left(\frac{\pi}{k+2}\right)} = \frac{\sin\left(\frac{(2j_2+1)}{k+2}\pi\right)}{\sin\left(\frac{\pi}{k+2}\right)}~.
\ee
This equation clearly only holds when $j_1=0$, that is for the parafermionic fields $\psi_{m_2}=\CO_{0,2m_2}$. However deformations by these fields can be ruled out by the $\mathbb Z_k$ symmetry, which requires
\be
[D_{\CO_{0,2m_1}}, \CO_{2j_2,2m_2}] = 0~, \ \ \ \forall m_1~.
\ee
These commutators lead to the condition
\be
 \frac{\sin\left(\frac{(2j_2+1)}{k+2}\pi\right)}{\sin\left(\frac{\pi}{k+2}\right)} \cdot e^{-\frac{4\pi i}{k}m_1m_2} = \frac{\sin\left(\frac{(2j_2+1)}{k+2}\pi\right)}{\sin\left(\frac{\pi}{k+2}\right)}\quad \Rightarrow \quad  e^{-\frac{4\pi i}{k}m_1m_2} = 1~,
\ee
which holds only for $m_2=0$ (but not for arbitrary $j_2$; however, we have already ruled out those fields).

Note that we can also prove that $[D_{\CO_{2j,0}}, \CO_{0,2m}] = 0$ using Theorem \ref{Thm2}. To do so, we note that 
\be
\sum_{c}N^{c}_{\psi_l,\psi_{k-l}} N^c_{\CO_{2j,0},\CO_{2\bar j,0}} = N^{\CO_{0,0}}_{\CO_{2j,0}\CO_{2\bar j,0}} = 1~.
\ee

To summarize, the commutator with the neutral fields rules out all of the perturbations except the ones generated by the parafermionic currents and the latter are ruled out by the commutation with group-like $\mathbb Z_k$ defects.\footnote{As discussed in the main text, one should only impose discrete symmetry defect commutators for symmetries that appear microscopically in the chain. For the case of higher-spin chains, this subtlety may potentially lead to realizations of the parafermionic CFTs having relevant deformations that are not protected by the $\mathbb{Z}_k$ discrete symmetries (these symmetries are then accidental symmetries of the continuum limit; see \cite{vernier2016elaborating} for a recent discussion).}

\end{appendices}

\newpage
\bibliography{chetdocbib}

\end{document}